\documentclass[reprint,aps,pre]{revtex4-2}
\usepackage{amsmath,amssymb,amsthm}
\usepackage[mathscr]{eucal}
\usepackage[colorlinks,linkcolor=blue,anchorcolor=green,citecolor=blue, urlcolor=blue,bookmarks=false,hypertexnames=true]{hyperref}
\usepackage{graphicx,lipsum}
\usepackage{epsf,epsfig,graphics}
\usepackage{orcidlink}
\usepackage{dcolumn}
\usepackage{bm}
\usepackage{subfigure}
\usepackage[section]{placeins}
\usepackage{float}
\usepackage{multirow}
\usepackage[margin=1.5cm]{geometry}
\usepackage[utf8]{inputenc}
\usepackage{xcolor, soul}
\sethlcolor{red}

\makeatletter
\newcommand\subsubsubsection[1]{%
  \paragraph{#1}\mbox{}\\
}
\makeatother

\begin{document}

\preprint{APS/123-QED}

\title{Exact diagonalization study of energy level statistics in harmonically confined interacting bosons}

\author{Mohd Talib \orcidlink{0000-0001-7261-3584}}
 \email{rs.mtalib@jmi.ac.in}
\author{M. A. H. Ahsan \orcidlink{0000-0002-9870-2769}}%
\email{mahsan@jmi.ac.in}
\affiliation{%
Department of Physics, Jamia Millia Islamia (A Central University), New Delhi 110025, India. }

\date{\today}

\begin{abstract}
We present an exact diagonalization study of the spectral properties of bosons harmonically confined in a quasi-2D plane and interacting via repulsive Gaussian potential. We consider the lowest $100$ energy levels for systems of $N=12, 16$ and $20$ bosons in two distinct regimes: (a) when the interaction energy is small compared to the trap energy (moderate interaction) and (b) when the interaction energy is comparable to the trap energy (strong interaction), for the non-rotating ($L_{z}=0$) as well as the rotating single-vortex state ($L_{z}=N$). For higher angular momenta, $L_{z}=2N$ and $L_{z}=3N$, only the strong interaction regime is considered. While the nearest-neighbor spacing distribution (NNSD) $P(s)$ and  the ratios of consecutive level spacings distribution $P(r)$ are used to study the short-range correlations, the Dyson-Mehta $\Delta_3$ statistic and the level number variance $\Sigma^2(L)$ are used to examine the long-range correlations. In the moderate interaction regime, the non-rotating system exhibits  Poisson distribution, a characteristic of the regular energy spectra. In the strong interaction regime, the non-rotating system exhibits chaotic behavior signified by  GOE distribution. Furthermore, in the rotating case for the single-vortex state ($L_{z} = N$) in the moderate interaction regime, the system exhibits signatures of weak chaos with some degree of regularity in the energy-level spectra. However, in the strong interaction regime for the rotating case with $L_{z} = N$, $2N$ and $3N$, the system exhibits strong chaotic behavior. The rotation is found to contribute to enhancement of chaotic behavior in the system for both the moderate and the strong interaction regimes. Our results of NNSD analysis are supported by the analysis of the ratios of consecutive level spacings distribution $P(r)$, which does not involve unfolding.
\end{abstract}

\maketitle

\section{Introduction}
The energy-level spectra of a system exhibits unique features intrinsic to the system. Random Matrix Theory (RMT) as a framework for analyzing the statistical properties of energy spectra in heavy nuclei was introduced by E. P. Wigner \cite{wigner55,wigner57,wigner58, wigner1967}. It was further developed by Mehta and Dyson \cite{mehta,dyson1, dyson2} and has since become an important tool in the statistical analysis of energy levels in complex quantum many-body systems. In order to compare various models with the predictions of RMT, it is essential to eliminate from the energy-level spectra, the system-specific properties. This procedure, commonly referred to in the literature as unfolding \cite{haake}, normalizes the mean level density to unity. While extracting system-specific features, the possible error arising out of the unfolding process can be minimized by fitting a low-degree polynomial or by performing linear fits over smaller intervals with moving average technique. Energy level statistics measures such as the nearest-neighbor spacing distribution (NNSD) $P(s)$ \cite{mehta, Guhr, Santos2010} and the distribution of the ratio of consecutive level spacings $P(r)$ \cite{huse} are used to probe short-range correlations, while the Dyson-Mehta $\Delta_3$ statistic \cite{mehta} and the level number variance $\Sigma^2(L)$ \cite{stockmann} provide insight into long-range correlations. The NNSD $P(s)$, the Dyson-Mehta $\Delta_3$ statistic and the level number variance $\Sigma^2(L)$ involve the unfolding of the energy-level spectra while in the distribution of the ratio of consecutive level spacings $P(r)$, the unfolding is not required. Consequently, the NNSD $P(s)$, the Dyson-Mehta $\Delta_3$ statistic and the level number variance $\Sigma^2(L)$ may be prone to errors. For this reason, the distribution of the ratio of consecutive level spacings $P(r)$ is more useful tool than the NNSD $P(s)$ for the short-range correlations. 

The Bohigas-Giannoni-Schmit (BGS) conjecture \cite{bohigas} states that the nearest-neighbor spacing distribution of the energy-level spectra of quantum systems whose classical counterpart is chaotic follows one of the random matrix ensembles: the Gaussian Orthogonal Ensemble (GOE) or the Gaussian Unitary Ensemble (GUE) or the Gaussian Symplectic Ensemble (GSE) as determined by the time-reversal symmetry of the Hamiltonian \cite{dyson1,dyson2}. On the other hand, the Berry-Tabor conjecture demonstrates that the quantum systems whose classical analog is integrable, the nearest-neighbor spacing distribution of energy spectra follows the Poisson statistics \cite{berry}. Both conjectures have received substantial theoretical and numerical support and serve as a foundation for quantum chaos. RMT has found applications across diverse fields, including the stock market \cite{plerou1, plerou2}, the atmospheric science \cite{santhanam} and the analysis of human electroencephalogram (EEG) data \cite{seba}, etc.  

The statistical properties of energy levels in interacting trapped bosons have been investigated for varying numbers of energy levels \cite{chakrabarti2012}, as well as for the lowest and the highest lying levels \cite{kamalika1, kamalika2, kamalika3}, for a given two-body interaction, using the approximate potential harmonic expansion method (PHEM). These works examine the effects of the two-body interaction and the trap energy on the energy level statistics, reporting deviation from the BGS conjecture. To the best of our knowledge, the effect of variation in the two-body interaction, the number of bosons and the rotation on the energy level statistics of trapped interacting bosons has not been explored. In the present study, we include rotation and extend the analysis to two distinct regimes: one in which the interaction energy is small compared to the trap energy (moderately interacting regime) and the other in which the interaction energy is comparable to the trap energy (strongly interacting regime). Our present study holds particular significance as the many-body energy levels obtained from the diagonalization of the Hamiltonian matrix are variationally exact. A recent experimental study on ultracold gas of erbium atoms has observed signatures of quantum chaos \cite{frisch2014}. Spectral statistics has also been explored across a range of systems to investigate the emergence of quantum chaos, including two-component interacting fermions in a double-well potential \cite{patrycja}, van der Waals clusters \cite{haldar2014}, Bose-Bose mixtures \cite{tran} and Hubbard model \cite{Kollath2010, marco, Mateos_2024, Haderlein_2024}. 

Following the experimental realization of Bose-Einstein condensation (BEC) in dilute atomic vapors of $^{87}$Rb \cite{anderson}, $^{23}$Na \cite{ketterle1995} and $^{39}$K \cite{Inguscio2001} with repulsive interaction, and $^{7}$Li with attractive interaction \cite{Hulet1995, Wilkin1998}, there has been a surge in research activity employing BEC systems as a testbed for quantum many-body theories. Notably, the creation of vortices in rotating BEC \cite{matthews, madison} has drawn significant attention from the superfluidity \cite{douglas,donnelly1991quantized, Pethick_Smith_2001} and superconductivity \cite{shaeer,Abrikosov} communities in condensed matter physics. The system of trapped interacting bosons with an externally impressed rotation, as described in Sec. \ref{1}, has been the subject of extensive theoretical and experimental investigations in the context of BEC. Statistical fluctuations in trapped interacting bosons are governed by two-body interactions characterized by the s-wave scattering length $a_{s}$ \cite{inouye1998,dalfovo1999}. In these systems there exist two energy scales, namely the trap energy ($\hbar \omega_{\perp}$) and the interaction energy ($\propto Na_{s}$, $N$ being the number of bosons). The interplay between these energy scales gives rise to rich physics within the system. There has been studies of various properties of the ground state such as the condensate fraction, breathing modes, vortex formation and vortex melting \cite{ahsan2001, Xia2001prl, Xia2001pra, imran3, imran4, imran5, hamid2022}. Our purpose of the current work is to investigate the statistical properties of the energy levels including the ground state and the low-lying excited states of the trapped interacting bosons, with focus on the interplay between the interaction energy and the trap energy.

The rest of the article is organized as follows. In Sec. \ref{1}, we present our many-body Hamiltonian for trapped bosons. Sec. \ref{backgroundtheory} describes the statistical tools employed in spectral analysis of the many-body system. In Sec. \ref{4}, we report our numerical results on the energy level statistics and conclude the work by outlining future direction in Sec. \ref{5}.

\section{\label{1}The Model Hamiltonian}
We start with a system of interacting bosons, each of mass $M$, trapped in a quasi-2D plane by a harmonic potential $V(r)=\frac{1}{2}M(\omega_{\perp}^2r_{\perp}^2+\omega_z^2z^2)$ with ${x}$-${y}$ symmetry. The system is also subjected to an externally impressed rotation with angular velocity $\Omega\equiv\Omega \hat{z}$. The many-body Hamiltonian for the system in the co-rotating frame is given by \cite{ahsan2001}
\begin{eqnarray*}
    \hat{H}^{\mathrm{rot}}=\hat{H}^{\mathrm{lab}}-\Omega.\hat{L}^{\mathrm{lab}}
\end{eqnarray*}
where\\
\begin{equation} \begin{split} \hat{H}^{\mathrm{lab}}=&\underbrace{\sum_{i=1}^{N}\left[\overbrace{\frac{1}{2M}\left(\frac{\hbar}{i}\nabla_{i}\right)^2+\frac{1}{2}M\omega_{\perp}^2(r_{\perp i}^2+\lambda_{z}^2z_{i}^2)}^{\hat{h}(\mathbf{r}_{i})}\right]}_{\hat{H}_{0}}\\ &+\frac{1}{2}\frac{4\pi \hbar^2 a_{sc}}{M}\left(\frac{1}{\sqrt{2\pi} \sigma} \right)^3\\ &\times \sum_{i\neq j}^{N}\exp^{-(\frac{1}{2\sigma^2})\{(r_{\perp i}-r_{\perp j})^2+(z_{i}-z_{j})^{2}\}} \end{split} \label{MBH} \end{equation}

\noindent where $\hat{H}_{0}=\sum_{i=1}^{N}\hat{h}(\mathbf{r}_{i})$ is the non-interacting part of the many-body Hamiltonian in the laboratory frame and\\ 

\begin{eqnarray*}
    \hat{L}_{z}^{\mathrm{lab}}=\sum_{i=1}^{N}\hat{l}_{z_{i}}^{\mathrm{lab}}=\frac{\hbar}{i}\sum_{i=1}^{N}(r_i\times\nabla_{i})_{z},
\end{eqnarray*}
$\hat{L}_{z}^{\mathrm{lab}}$ is the total angular momentum of the system in the laboratory frame. Here $r_{\perp}=\sqrt{x^2+y^2}$ is the radial distance of the particle from the trap axis; $\omega_{\perp}$ and $\omega_z$ are the confining harmonic frequencies in the $x$-$y$ plane and the axial direction, respectively. The first two terms of the many-body Hamiltonian in Eq. (\ref{MBH}) comprises $\hat{H}_{0}$ and represent the kinetic energy and the external confining potential, while the last term describes the two-body interaction, modelled by a potential Gaussian in particle-particle separation. We introduce the anisotropic parameter $\lambda_{z}\equiv \frac{\omega_z}{\omega_{\perp}}\gg$1 which signifies that the trap is highly oblate spheroidal and hence the confined system is effectively quasi-2D. We choose $\hbar \omega_{\perp}$ as the unit of energy and define $a_{\perp}$=$\sqrt{\frac{\hbar}{M\omega_{\perp}}}$ as the unit of length. In the Gaussian potential, $\sigma$ (scaled by $a_{\perp}$) is the interaction range of the potential. The effective two-body interaction strength in 2D, after tracing out the $z$-coordinate becomes
\begin{equation*}
 g_{2} = \frac{4\pi \hbar^{2} a_{sc}}{M} \, \frac{1}{\sqrt{2\pi}} \sqrt{\frac{\lambda_{z}}{a_{\perp}^{2}\left[1 + \left(\tfrac{\sigma}{a_{\perp}}\right)^{2} \lambda_{z}\right]}},   
\end{equation*}
which in the limit $\frac{\sigma}{a_{\perp}}\ll 1$ and in the dimensionless form reduces to 
\begin{equation}
  g_{2}=\frac{4\pi a_{sc}}{a_{\perp}} \sqrt{\frac{\lambda_{z}}{2\pi}}
\end{equation}
 where $a_{sc}$ is the s-wave scattering length for the two-body interaction. Experimentally, one can vary the scattering length $a_{sc}$ employing Feshbach resonance \cite{inouye1998,chin2010}. In this study we assume $a_{sc} > 0$ so that the effective interparticle potential is in the repulsive regime. When the interaction range approaches zero i.e $\sigma \to 0 $, the Gaussian interaction reduces to the $\delta$-function potential $V(r_{\perp i},r_{\perp j})=g_{2}\delta(r_{\perp i}-r_{\perp j})$ \cite{dalfovo1999}.

The eigenspectrum for the many-body Hamiltonian $\hat{H}^{\mathrm{lab}}$ in Eq. (\ref{MBH}), in given subspaces of total angular momentum $\hat{L}_{z}^{\mathrm{lab}}$, in the laboratory frame are obtained variationally by setting up the $N$-body eigenvalue equation
\begin{equation}\label{N-body}
    \hat{H}^{\mathrm{lab}}\Psi(\mathbf{r}_{1}, \mathbf{r}_{2}, \ldots, \mathbf{r}_{N})=E\Psi(\mathbf{r}_{1}, \mathbf{r}_{2}, \ldots, \mathbf{r}_{N}),
\end{equation}
for the Hamiltonian in Eq. (\ref{MBH}) with
 \begin{equation*}
     \Psi(\mathbf{r}_{1}, \mathbf{r}_{2}, \ldots, \mathbf{r}_{N})=\sum_{\nu}C_{\nu}\phi_{\nu}(\mathbf{r}_{1}, \mathbf{r}_{2}, \ldots, \mathbf{r}_{N}),
 \end{equation*}
where the set of parameters $\{C_{\nu}\}$ are to be determined variationally and the set $\{\phi_{\nu}(\mathbf{r}_{1}, \mathbf{r}_{2}, \ldots, \mathbf{r}_{N})\}$ are the $N$-body basis functions, satisfying the eigenvalue equation $\hat{H}_{0}\phi_{\nu}=E_{\nu}\phi_{\nu}$, where $H_{0}$ is the non-interacting part of the Hamiltonian in Eq. (\ref{MBH}) and $\{\nu\}$ is the set of $N$-body quantum numbers. The $\phi_{\nu}$ is constructed as the symmetrized product of the harmonic oscillator single-particle basis functions in 2D satisfying the eigenvalue equation $\hat{h}(\mathbf{r})u_{n,m} (\mathbf{r})=\epsilon_{n,m}u_{n,m}(\mathbf{r})$ 
with 

\begin{align*}
	\epsilon_{n,m} &= \left(\underbrace{2n_{r} + |m|}_{n} + 1\right)\hbar\omega_{\perp}
	\equiv (n + 1)\hbar\omega_{\perp}, \label{n} \\[4pt]
	&\quad n = 0, 1, 2, \ldots, \quad 
	m = \underbrace{-n, -n + 2, \ldots, n - 2, +n}_{\text{in steps of 2}}. \nonumber
\end{align*}
and 
\begin{equation*}
\begin{aligned}
	u_{n,m}(r_{\perp}, \phi) =& 
	\sqrt{\frac{\alpha_{\perp}^{2}}{\pi}
		\frac{\left(\tfrac{1}{2}[n - |m|]\right)!}{\left(\tfrac{1}{2}[n + |m|]\right)!}}
	\, e^{-\frac{\alpha_{\perp}^{2} r_{\perp}^{2}}{2}}
	\, e^{i m \phi}
	(\alpha_{\perp} r_{\perp})^{|m|}\\
	&\times L_{\tfrac{1}{2}(n - |m|)}^{|m|}(\alpha_{\perp}^{2} r_{\perp}^{2}),
	\quad \alpha_{\perp} = \sqrt{\frac{M \omega_{\perp}}{\hbar}},
\end{aligned}
\end{equation*}

\begin{equation*}
	\hat{l}_{z}\,u_{n,m}(r_{\perp}, \phi) = m \hbar \, u_{n,m}(r_{\perp}, \phi).
\end{equation*}

On carrying out the variational scheme, the $N$-body eigenvalue problem in Eq. (\ref{N-body}) reduces to the matrix eigenvalue problem $\mathbf{H}\mathbf{C}=E\mathbf{C}$, where $\mathbf{C}$ is the column vector comprising of the variational parameters $\{C_{\nu}\}$ and $\mathbf{H}$ is the matrix representation of the Hamiltonian $\hat{H}^{\mathrm{lab}}$ in Eq. (\ref{MBH}). The Hamiltonian matrix $\mathbf{H}$ is large but sparse and the lowest $100$ eigenvalues and eigenvectors are determined through iterative diagonalization \cite{DAVIDSON197587}. The detailed scheme for finding the eigenspectrum can be found in Ref. \cite{ahsan2001}. 

Having obtained the eigenspectrum for the Hamitonian in Eq. (\ref{MBH}) in given subspaces of the total angular momentum, we proceed to carry out the statistical analysis of the energy levels as described in the following.

\section{\label{backgroundtheory}Statistical tools for spectral analysis}
In the following, we outline the statistical tools employed in our analysis of energy-levels both for the short-range and the long-range correlations.

\subsection{\label{NNSDAnalysis}Nearest-neighbor spacing distribution}
The histogram of nearest-neighbor spacing distribution (NNSD) $P(s)$ is used to study the short-range correlation properties of the system \cite{mehta, Guhr, reichl}. Prior to the statistical analysis of the spectra, we need to separate out the smooth part of the density of states from the fluctuating part using the unfolding procedure. In this work we unfold the spectra using a polynomial of degree 6.
We calculate the nearest-neighbor spacing as $s=\epsilon_{n+1}-\epsilon_{n}$ and determine the probability distribution of spacing $P(s)$. We compare the histogram of NNSD $P(s)$ with the theoretical distributions of Poisson and GOE. The integrable spectra obeys the Poisson distribution $P(s)=e^{-s}$ while the nonintegrable spectra obeys the GOE distribution $P(s)=\frac{\pi}{2}se^{-\frac{\pi s^2}{4}}$. Further, we compare the Poisson and the GOE distribution with the Brody distribution given by \cite{brody}
\begin{equation}
    P_{b}(s)=(1+b)as^{b}{\exp(-as^{1+b})}   \label{fittingbrody}
\end{equation}
where $b$ is the Brody parameter and $a=[\Gamma(\frac{2+b}{1+b})]^{1+b}$ [$\Gamma$ is Euler's gamma function]. The parameter $b$ is like an interpolation parameter which for $b=0$ corresponds to the Poisson distribution while $b=1$ corresponds to the GOE distribution.

Level repulsion between the energy-levels described by $P(s) \propto s^{\beta}$ with $\beta = 1$ corresponds to systems that follow the GOE distribution. This repulsion is a characteristic of chaotic systems. For small values of the level spacing $s$, the distribution $P(s)$ approaches zero, i.e., $P(s) \approx 0$ as $s \to 0$, indicating the presence of level repulsion at short energy spacings \cite{rosenzweig}.

\subsection{\label{mehta}The Dyson-Mehta $\Delta_3(L)$ statistic}
To study the long-range correlations in the system, the Dyson-Mehta $\Delta_3(L)$ statistic is employed \cite{mehta}. For the chosen energy interval $[a,a+L]$ of length $L$, the $\Delta_3$ statistic is defined by
\begin{equation}
\Delta_3(a;L)= \frac{1}{L}\underset{A,B}{\min}\int_{a}^{a+L}[N(\epsilon)-A\epsilon-B]^2d\epsilon.
\end{equation}
It is the least-square deviation of the straight line $A\epsilon+B$ from the step function $N(\epsilon)$. By averaging $\Delta_3(a;L)$ over the chosen energy intervals, we obtain the spectral average $\Delta_3(L)$, which quantifies the deviation of the unfolded spectrum from a uniformly spaced spectrum. From now on, the spectral average $\langle \Delta_3(L)\rangle$ will be abbreviated as $\Delta_3(L)$. For the Poisson distribution, the spectral average $\Delta_3(L)$ varies linearly with $L$ as
\begin{equation*}
    \Delta_3(L)=\frac{L}{15}
\end{equation*}
while for the GOE distribution, the spectral average $\Delta_3(L)$ varies logarithmically \cite{drod} as
\begin{equation*}
     \Delta_3(L)=\frac{1}{\pi^2}\ln(L)-0.00695.
\end{equation*}

\subsection{\label{level} Level number variance ${\Sigma^2(L)}$}
The level number variance $\Sigma^2(L)$ is statistical measure used to characterize the long-range correlations in complex quantum systems \cite{stockmann, Guhr} and 
is defined as

\begin{equation}
    {\Sigma^2(L)}=\langle(N(\epsilon,L)-\langle N(\epsilon,L) \rangle)^2\rangle.  \label{lnveqn}
\end{equation}
Simplifying Eq. (\ref{lnveqn}), we obtain
\begin{equation}
    {\Sigma^2(L)}=\langle N(\epsilon,L)^2\rangle-\langle N(\epsilon,L)\rangle^2.
\end{equation}
For the Poissonian distribution, the level number variance varies linearly with $L$ as
\begin{equation*}
   \Sigma^2(L)=L 
\end{equation*}
while for systems obeying the GOE distribution, it varies logarithmically as
\begin{equation*}
   \Sigma^2(L)=\frac{2}{\pi^2}[\log(2 \pi L)+1.577215-\frac{\pi^2}{8}] . 
\end{equation*}
The $\Delta_3(L)$ statistic and the level number variance $\Sigma^2(L)$ are related through the equation
\begin{equation}
   \Delta_3(L)=\frac{2}{L^4}\int_{0}^{L}(L^3-2L^2\epsilon+\epsilon^3) \Sigma^2(\epsilon)d\epsilon.    \label{deleqn}
\end{equation}
From Eq. (\ref{deleqn}), we observe that $\Delta_3(L)$ is more uniform than $\Sigma^2(L)$, as it is the integrated version of the level number variance. This uniformity is evident in the plots of the $\Delta_3(L)$ statistic and the level number variance $\Sigma^2(L)$.

\subsection{Distribution of the ratio of consecutive level spacings}

The NNSD $P(s)$, the Dyson-Mehta $\Delta_{3}$ statistic and the level number variance $\Sigma^2(L)$, discussed in Sections \ref{NNSDAnalysis}, \ref{mehta} and \ref{level}, respectively, depend strongly on the unfolding procedure. Apart from systems where the expression for the mean density of states is known exactly, the unfolding of the energy spectrum introduces error. To circumvent the unfolding procedure, Oganesyan and Huse introduced \cite{huse} gap-ratio $r$ defined as 

\begin{equation}
    r_n=\frac{\min(s_n,s_{n-1})}{\max(s_n,s_{n-1})}
\end{equation}
where $s_n=E_n-E_{n-1}$.\\

To determine the distribution of the ratio of consecutive level spacings $P(r)$, we make the histogram of the ratios of consecutive level spacings. The histogram obtained from the numerical data are compared with the theoretical distributions of Poisson and GOE given by \cite{atas}

\begin{equation}
    P_{\mathrm{Poisson}}(r)=\frac{2}{(1+r)^2}   \label{eqP} 
\end{equation}
\begin{equation}
    P_{\mathrm{GOE}}(r)=\frac{27}{4}\frac{r+r^2}{(1+r+r^2)^{5/2}}.  \label{eqW}
\end{equation}

The advantage of the distribution of the ratios of consecutive level spacings $P(r)$ over the spacing distribution $P(s)$ is that we do not have to unfold the spectra. Thus, the distribution $P(r)$ is more useful quantity for comparison with experiments than the distribution of level spacings $P(s)$. The distribution $P(r)$ is used in many contexts of quantum many-body physics, such as in the investigation of many-body localization \cite{huse,iyer2013, pal2010,emilio2012}, Bose-Hubbard models \cite{Kollath2010, collura2012} and the interaction driven metal-insulator crossover in the Hubbard model \cite{Equbal_2025}. The average value of the gap ratio for the Poisson distribution is $\langle r\rangle_{\mathrm{Poisson}}=2\ln(2)-1 \approx 0.386$ \cite{atas}, and for the GOE distribution, $\langle r\rangle_{\mathrm{GOE}} \approx 0.530$ \cite{atas}. These averages of gap ratios can be obtained using the distributions given in Eqs. (\ref{eqP}) and (\ref{eqW}).

\section{\label{4}Results \& Discussions}
We consider $N=12, 16$ and $20$ bosonic atoms of $^{87}$Rb confined in a quasi-2D harmonic trap, interacting via a repulsive Gaussian potential with system parameters described as in the following \cite{dalfovo,baym}. For the quasi-2D harmonic trap, we take the anisotropy parameter as $\lambda_{z}=\frac{\omega_{z}}{\omega_{\perp}}=4$. The axial frequency is taken to be $\omega_{z}=2\pi \times 220$ $Hz$ leading to the trap length $a_{\perp}=\sqrt{\frac{\hbar}{M\omega_{\perp}}}=1.446$ $\mu m$ where $M$ is the mass of the $^{87}$Rb atom. The interaction range in the Gaussian potential is taken as $\sigma$= 0.1$a_{\perp}$. The relevant experimental parameter for the two-body interaction strength in the mean-field approximation is $Na_{sc}/a_{\perp}$ for the contact potential \cite{dalfovo1999}. In our exact diagonalization study, $N$ is to be limited to a few tens of particles due to exponentially increasing dimensionality of the many-particle Hilbert space. To achieve the value of the parameter $Na_{sc}/a_{\perp}$ relevant to experimental situations, we parametrically increase the value of the $s$-wave scattering length with $a_{sc}=1000a_{0}$ for moderate interaction regime (when the interaction energy is small compared to the trap energy) and  $a_{sc}=10000a_{0}$ for strong interaction regime (when the interaction energy is comparable to the trap energy) where $a_{0}=0.05292$ $nm$ is the Bohr radius. With these parameter values, the value of the dimensionless interaction strength parameter $g_{2}=\frac{4\pi a_{sc}}{a_{\perp}} \sqrt{\frac{\lambda_{z}}{2\pi}}$ for the quasi-2D system turns out to be $g_{2}=0.3669$ for moderately interacting regime and $g_{2}=3.669$ for strongly interacting regime. 

\begin{figure*}[t]
%\begin{center}
\centering 

    \subfigure[$N=12$]
   {\label{fig:12brody}\includegraphics[width=0.32\linewidth]{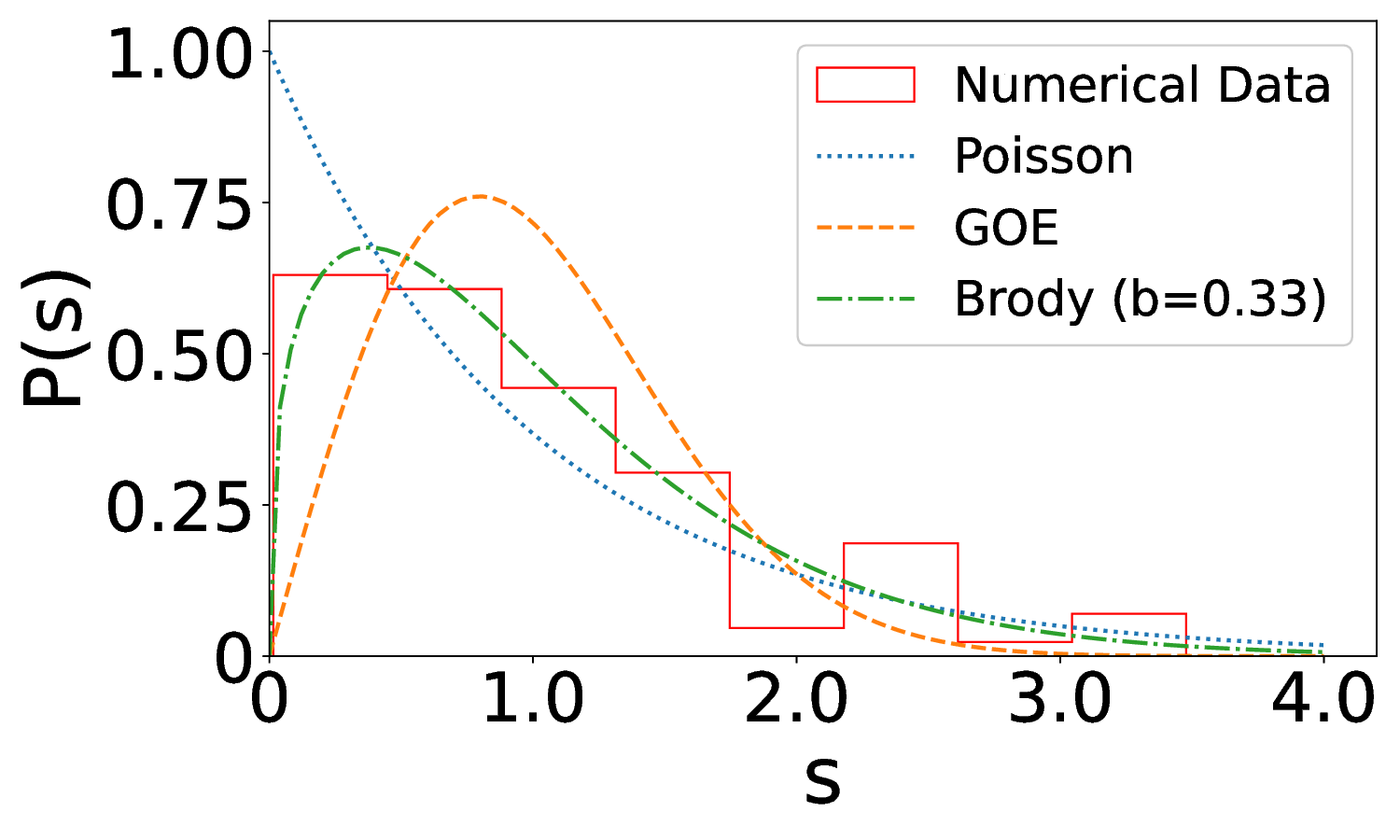}} 
    \subfigure[$N=16$]
    {\label{fig:16brody}\includegraphics[width=0.32\linewidth]{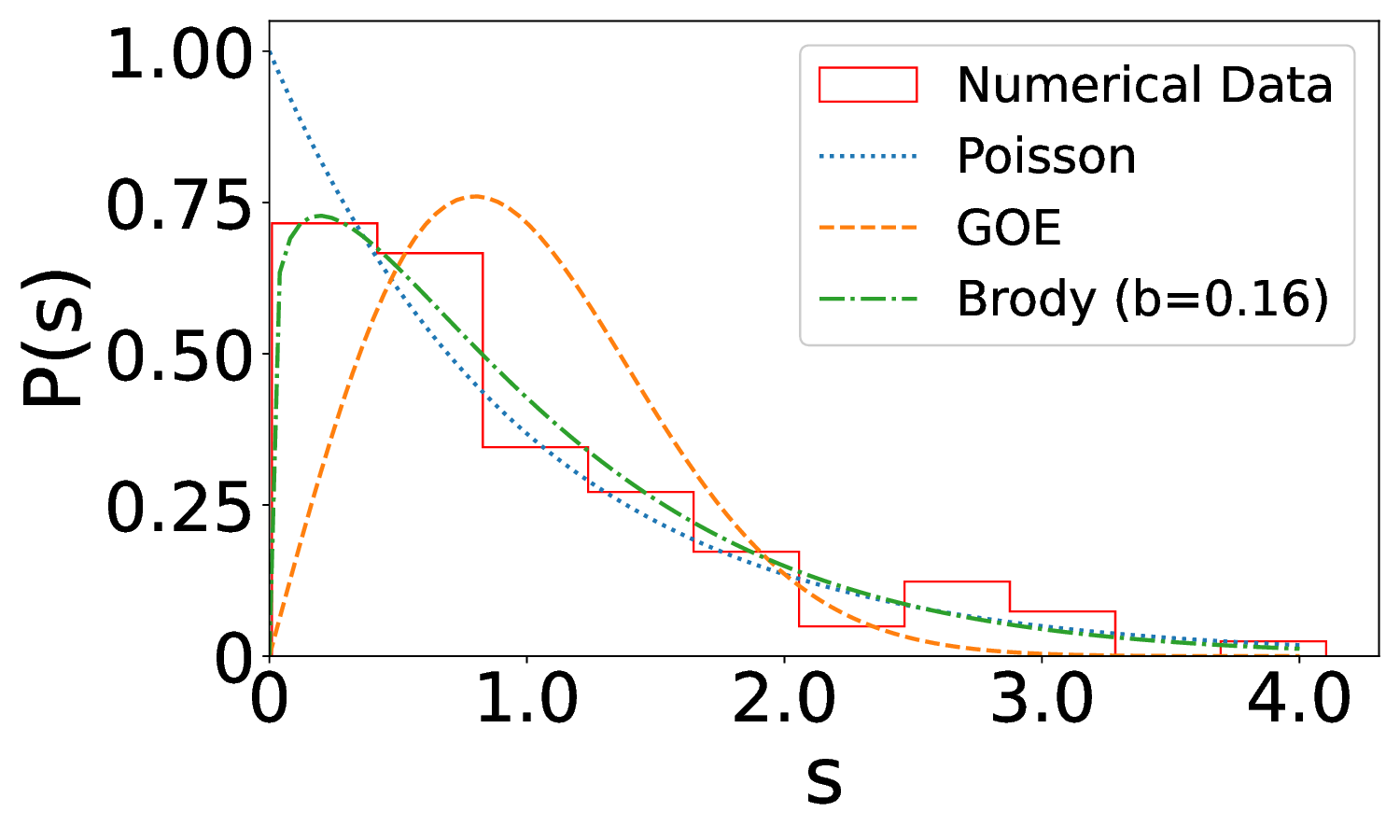}}
     \subfigure[$N=20$]{\label{fig:20brody}\includegraphics[width=0.32\linewidth]{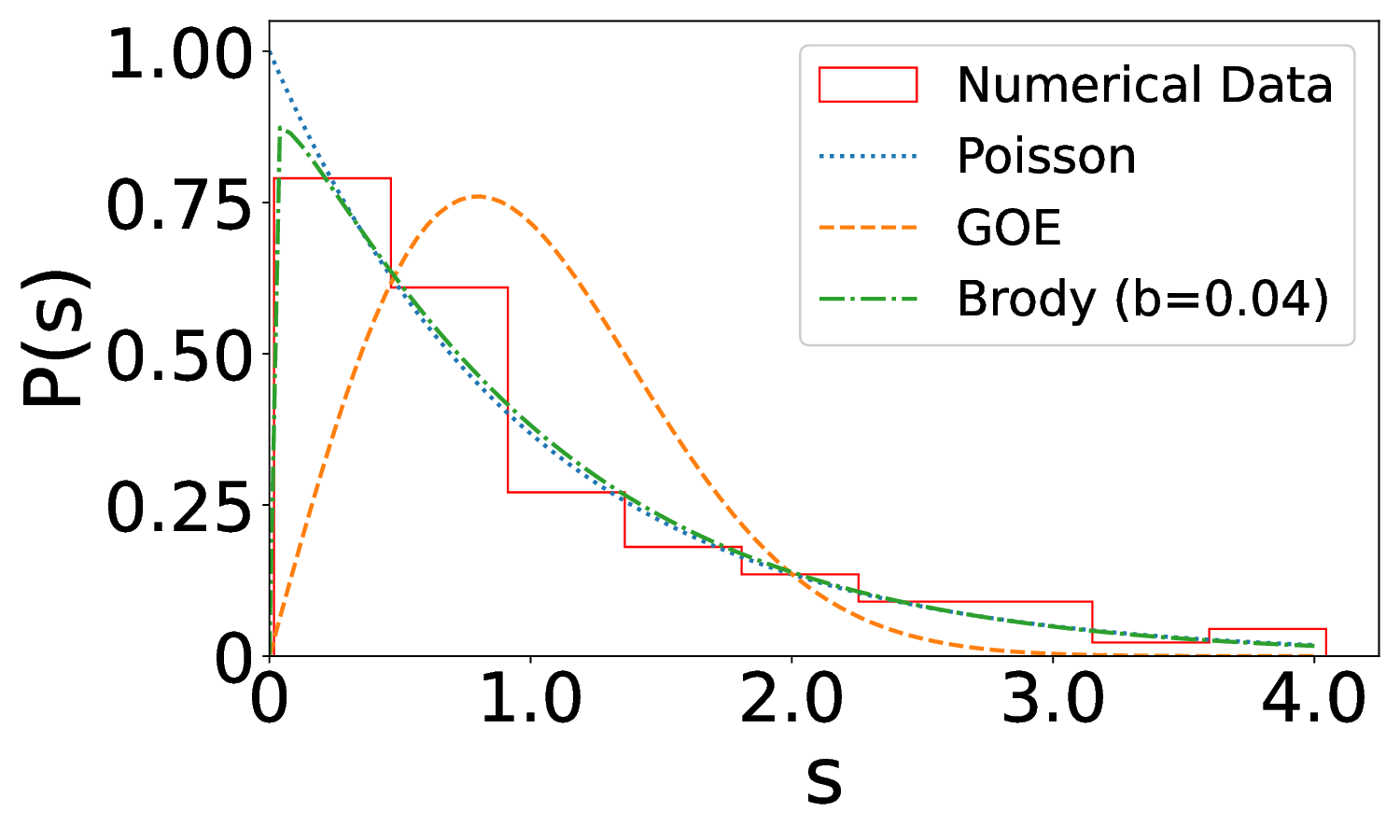}}
      \subfigure[$N=12$]{\label{fig:12gap}\includegraphics[width=0.32\linewidth]{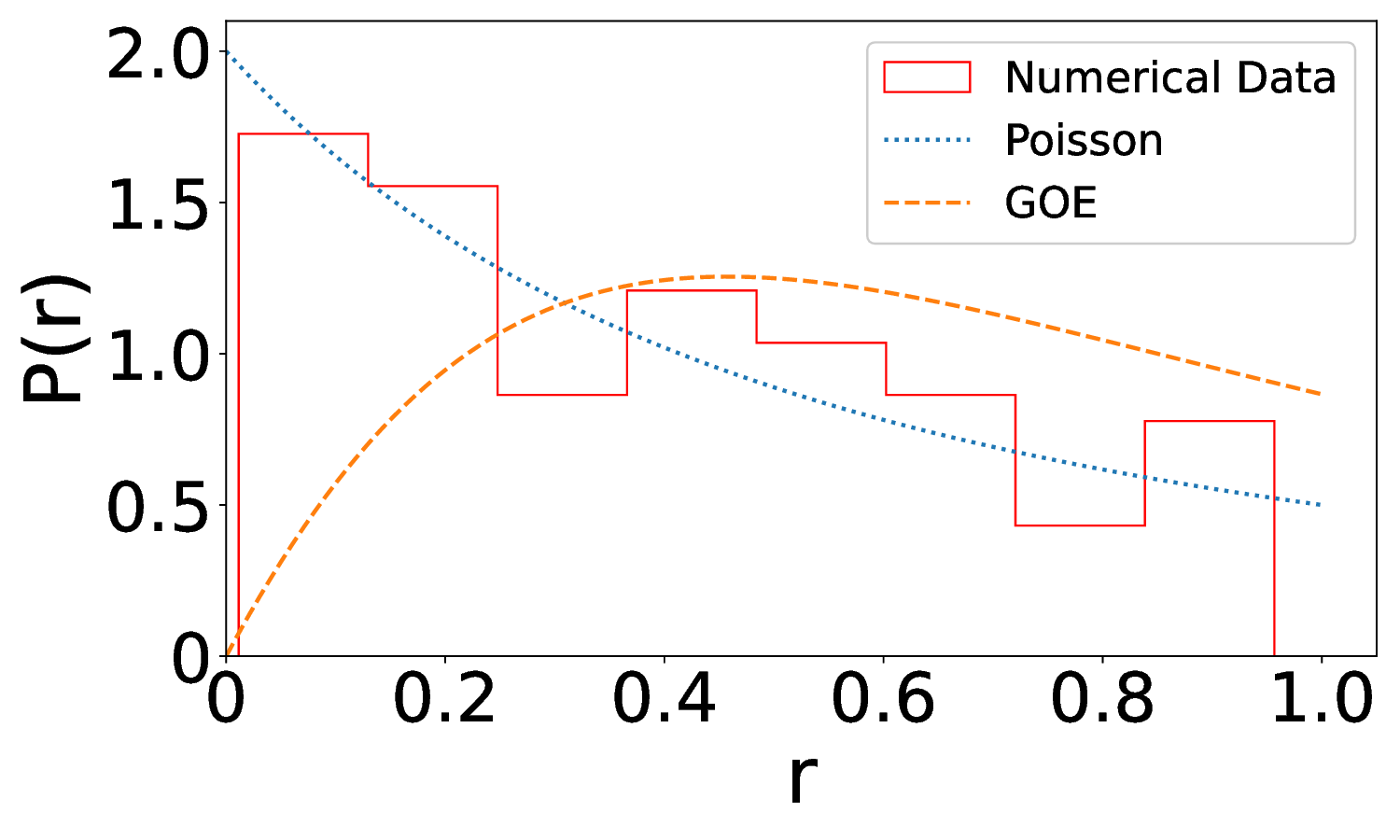}} 
     \subfigure[$N=16$]{\label{fig:16gap}\includegraphics[width=0.32\linewidth]{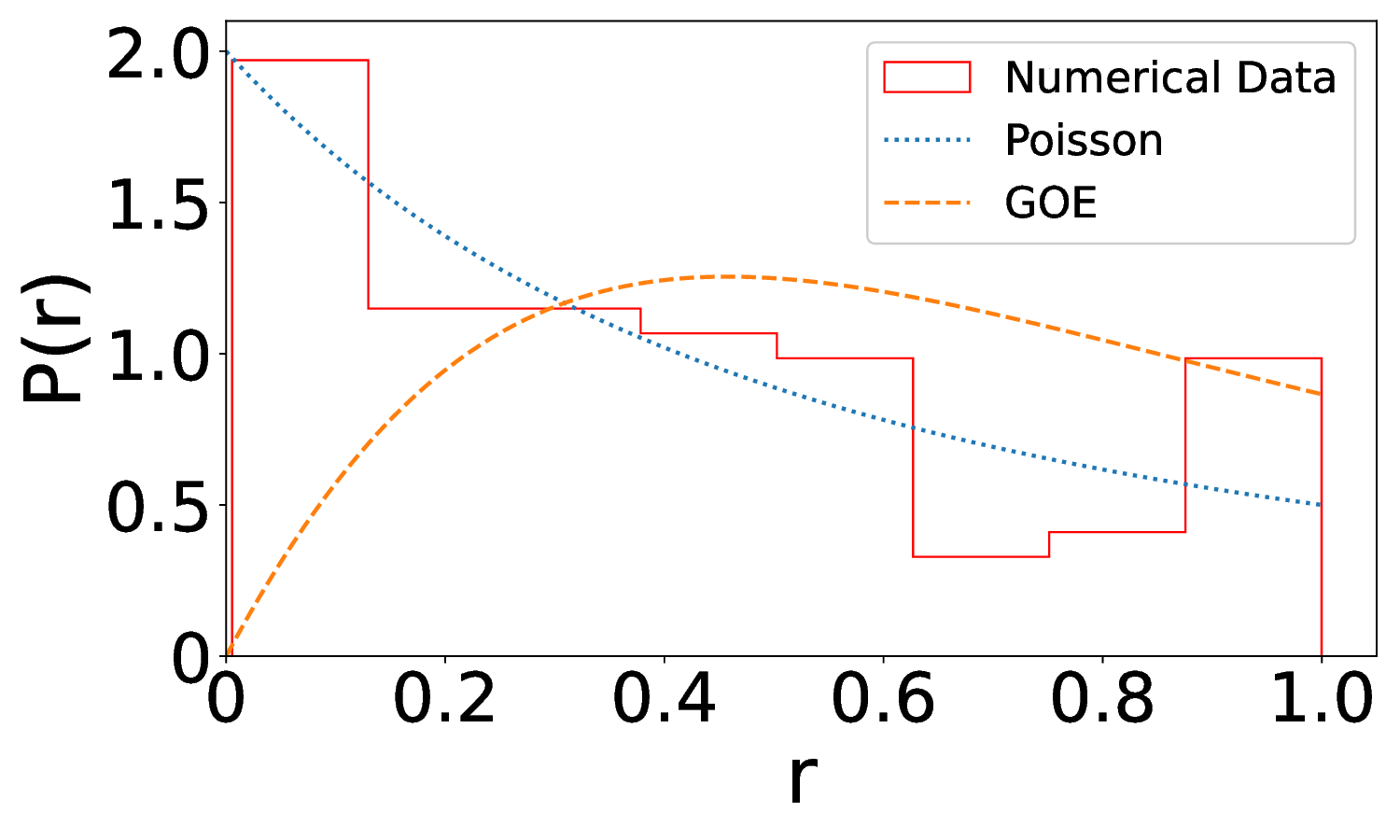}}
    \subfigure[$N=20$]{\label{fig:20gap}\includegraphics[width=0.32\linewidth]{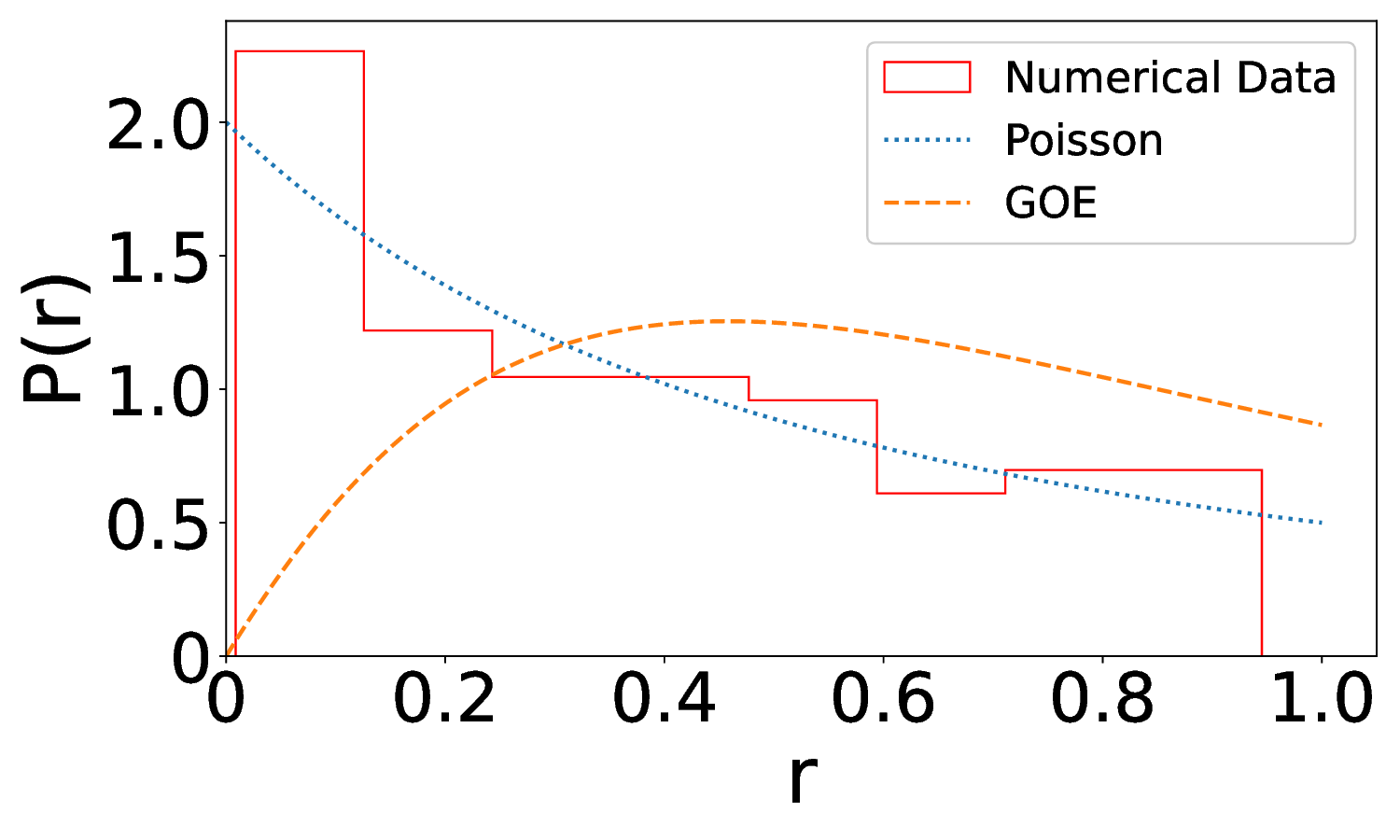}}
     
    \caption{(Color online) The nearest-neighbor spacing distribution $P(s)$ (upper panel) and the distribution of the ratio of consecutive level spacings $P(r)$ (lower panel) in the moderate interaction regime with $g_{2}=0.3669$ for $N=12, 16$ and $20$ bosons with total angular momentum $L_{z}=0$. The histogram in each graphs represents our numerical result for the lowest 100 energy levels. The blue dotted curve corresponds to the Poisson distribution, the orange dashed curve to the GOE distribution and the green dash-dotted curve to the Brody distribution with fitting parameter $b$.}
    \label{fig:NNSDnonrotmoderate}
\end{figure*}

We now present our numerical results on the spectral analysis employing the lowest $100$ energy levels for the non-rotating case in Sec. \ref{NRCase} and for the rotating case in Sec. \ref{Rcase}, considering both the moderate and the strong interaction regimes. To study the effect of the number of bosons $N$ on the statistical behavior of the energy levels, we vary the number of bosons as $N=12, 16$, $20$. Our aim is to understand how the variation in two-body interaction, rotation and number of bosons affect the level statistics and hence the spectral correlations.

\begin{table}[t]
	\caption{Fitted values of the Brody parameter $b$ in the non-rotating case for different number of bosons  in the moderate ($g_{2}=0.3669$) and the strong ($g_{2}=3.669$) interaction regimes.}
    \label{tablebrody1}
	\centering
	\begin{tabular}{|c|c|c|c|} \hline
		& Number of bosons, $N$ & \multicolumn{2}{c|}{Brody parameter $b$} \\ \cline{3-4}
		& & $g_{2}=0.3669$ & $g_{2}=3.669$ \\ \hline
		& 12 & 0.33 & 0.53 \\ \hline
		& 16 & 0.16 & 0.85 \\ \hline
		& 20 & 0.04 & 0.51 \\ \hline
		Poisson &  & \multicolumn{2}{c|}{0} \\ \hline
		GOE     &  & \multicolumn{2}{c|}{1} \\ \hline
	\end{tabular}
\end{table}

\subsection{\label{NRCase} Non-rotating case ($L_{z}=0$)}
To gain insight into the statistics of the energy levels, we investigate the short-range and the long-range correlations. 

\begin{figure*}[t]
\centering
    
     \subfigure[$N=12$]{\label{fig:12brody9.151}\includegraphics[width=0.32\linewidth]{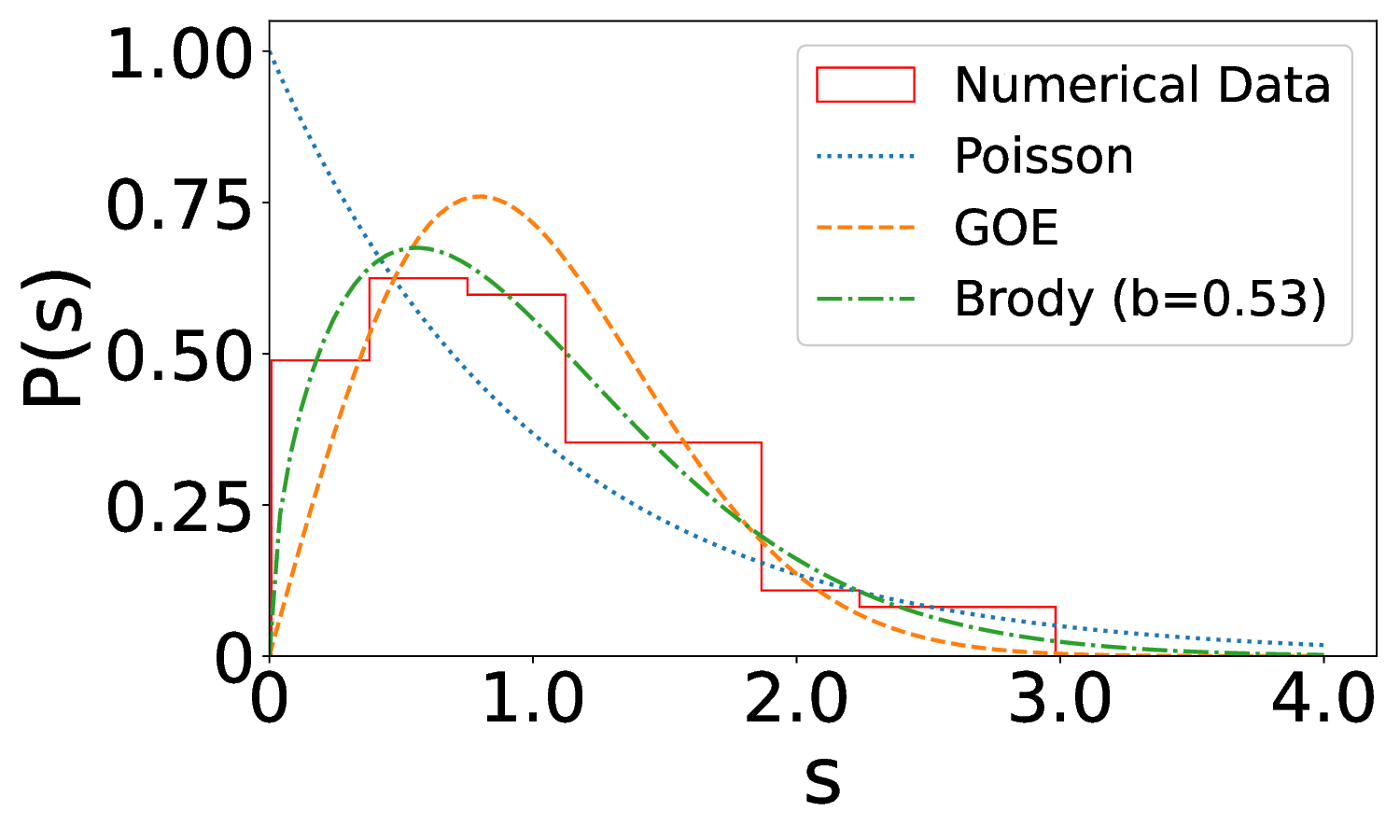}} 
     \subfigure[$N=16$]{\label{fig:16brody9.151}\includegraphics[width=0.32\linewidth]{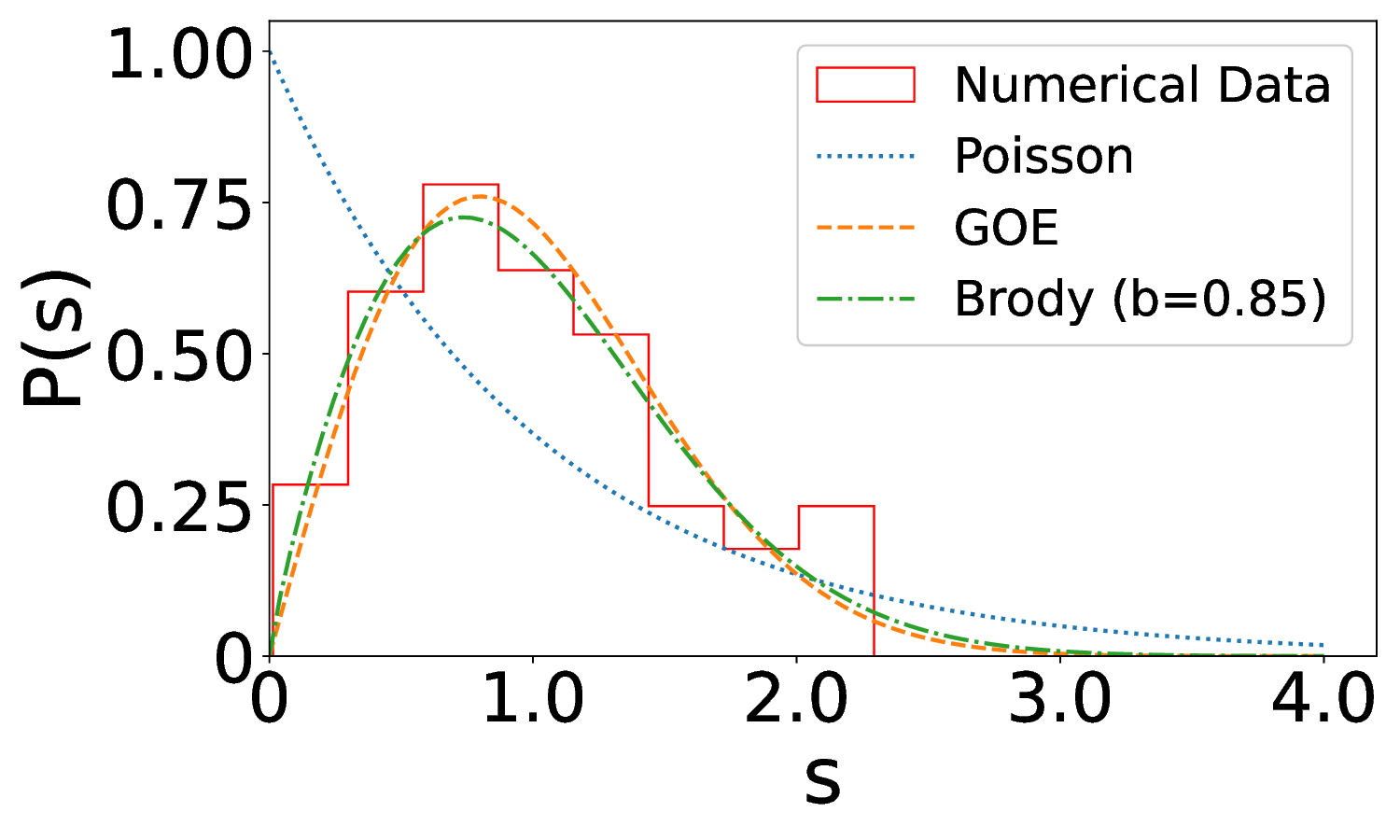}}
      \subfigure[$N=20$]{\label{fig:20brody9.151}\includegraphics[width=0.32\linewidth]{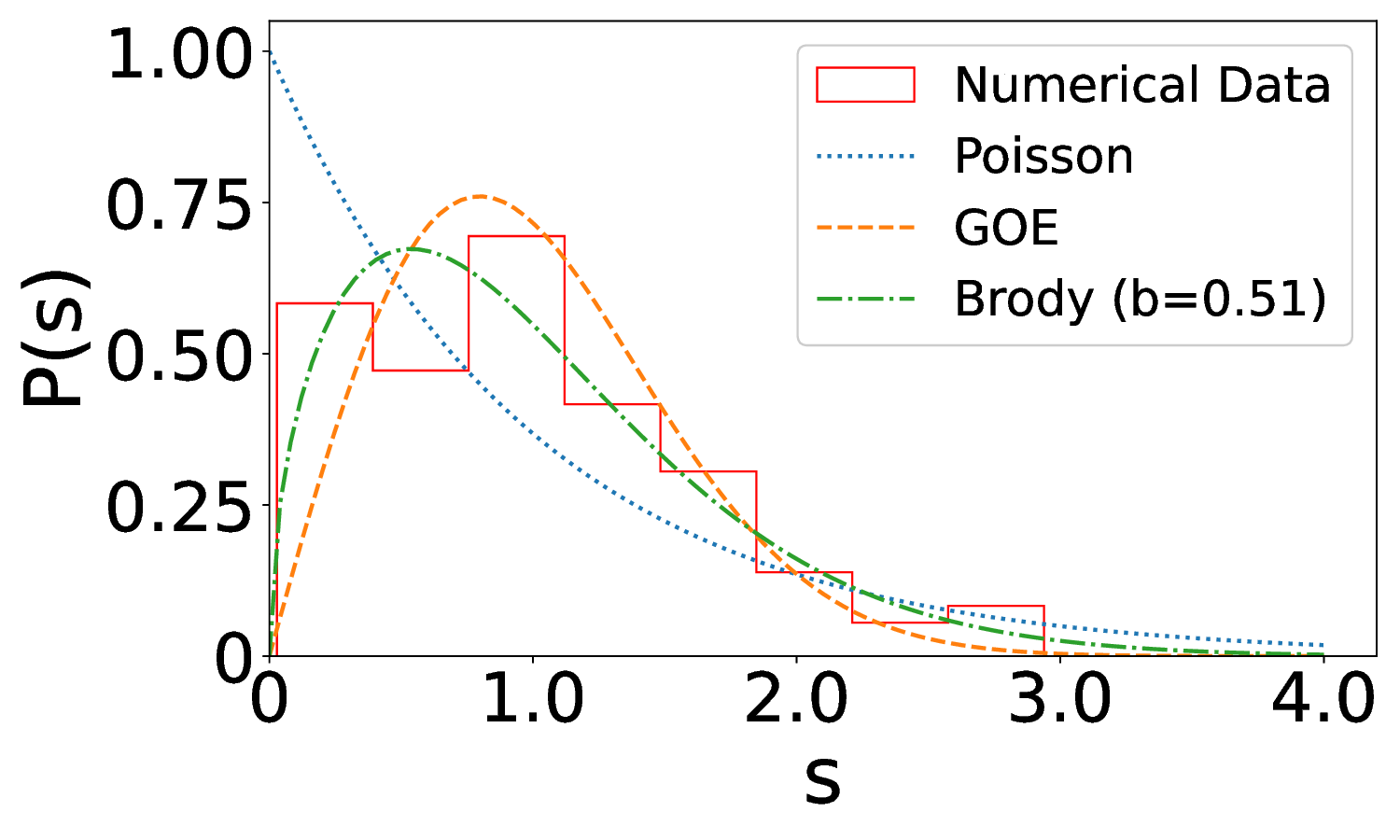}}
       \subfigure[$N=12$]{\label{fig:12gap9.151}\includegraphics[width=0.32\linewidth]{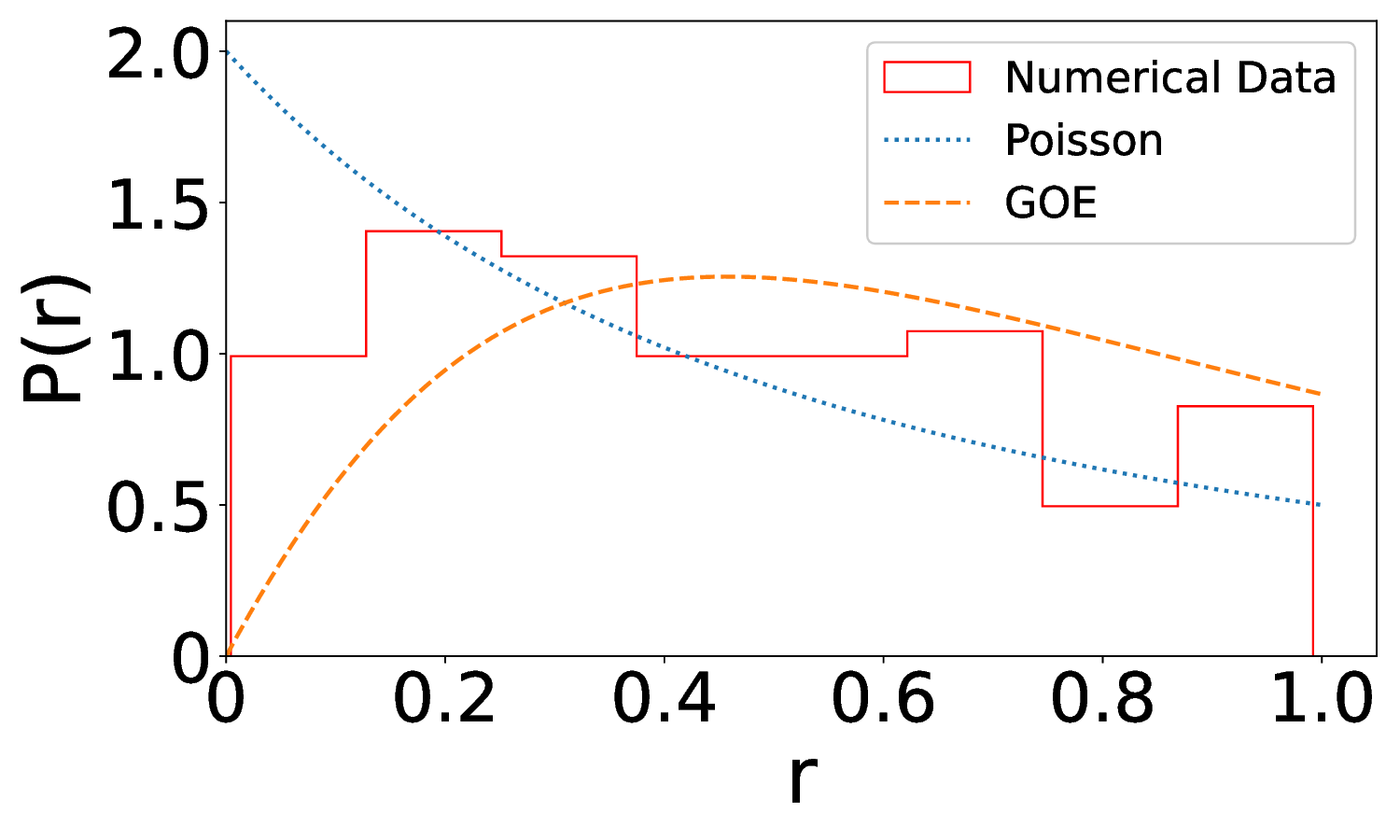}} 
     \subfigure[$N=16$]{\label{fig:16gap9.151}\includegraphics[width=0.32\linewidth]{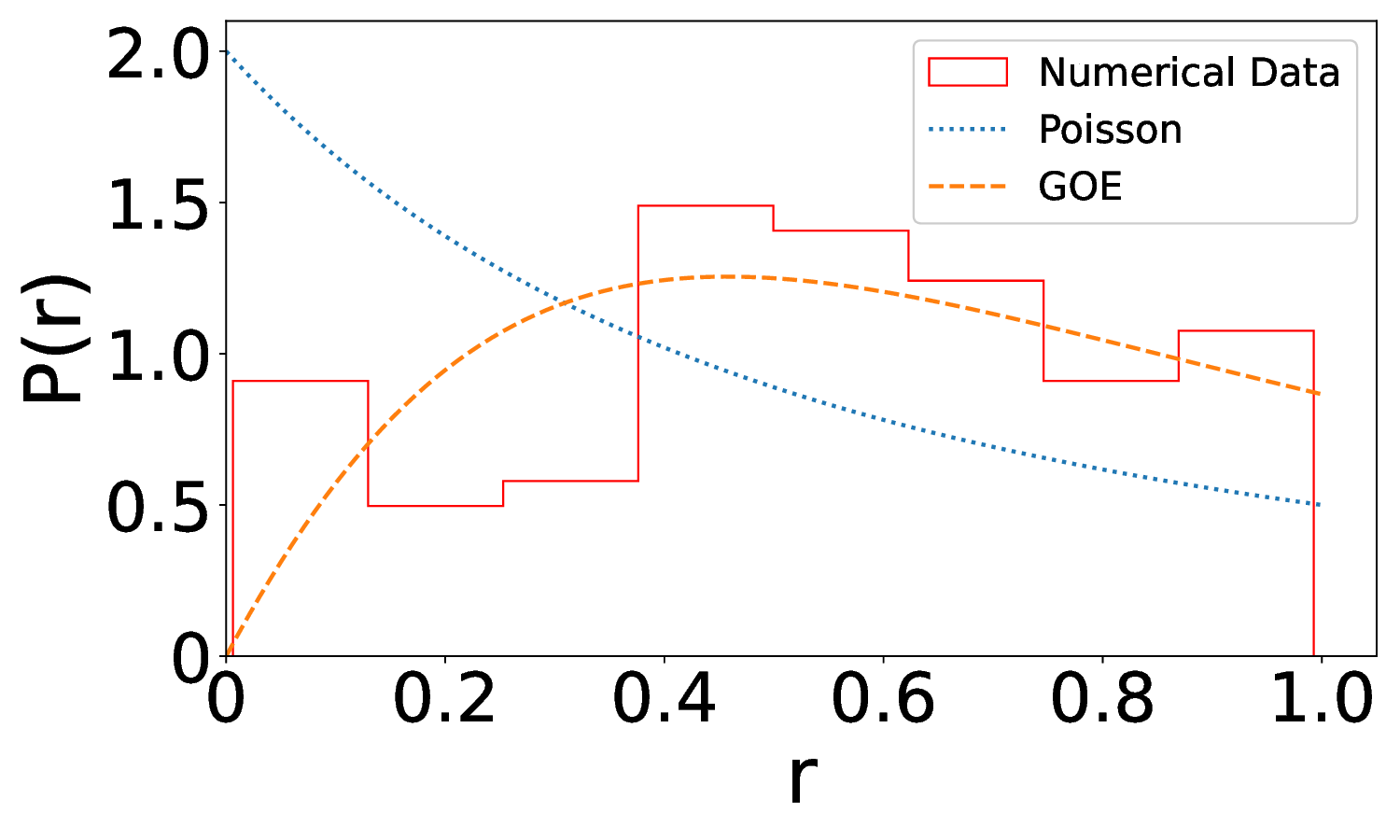}}
      \subfigure[$N=20$]{\label{fig:20gap9.151}\includegraphics[width=0.32\linewidth]{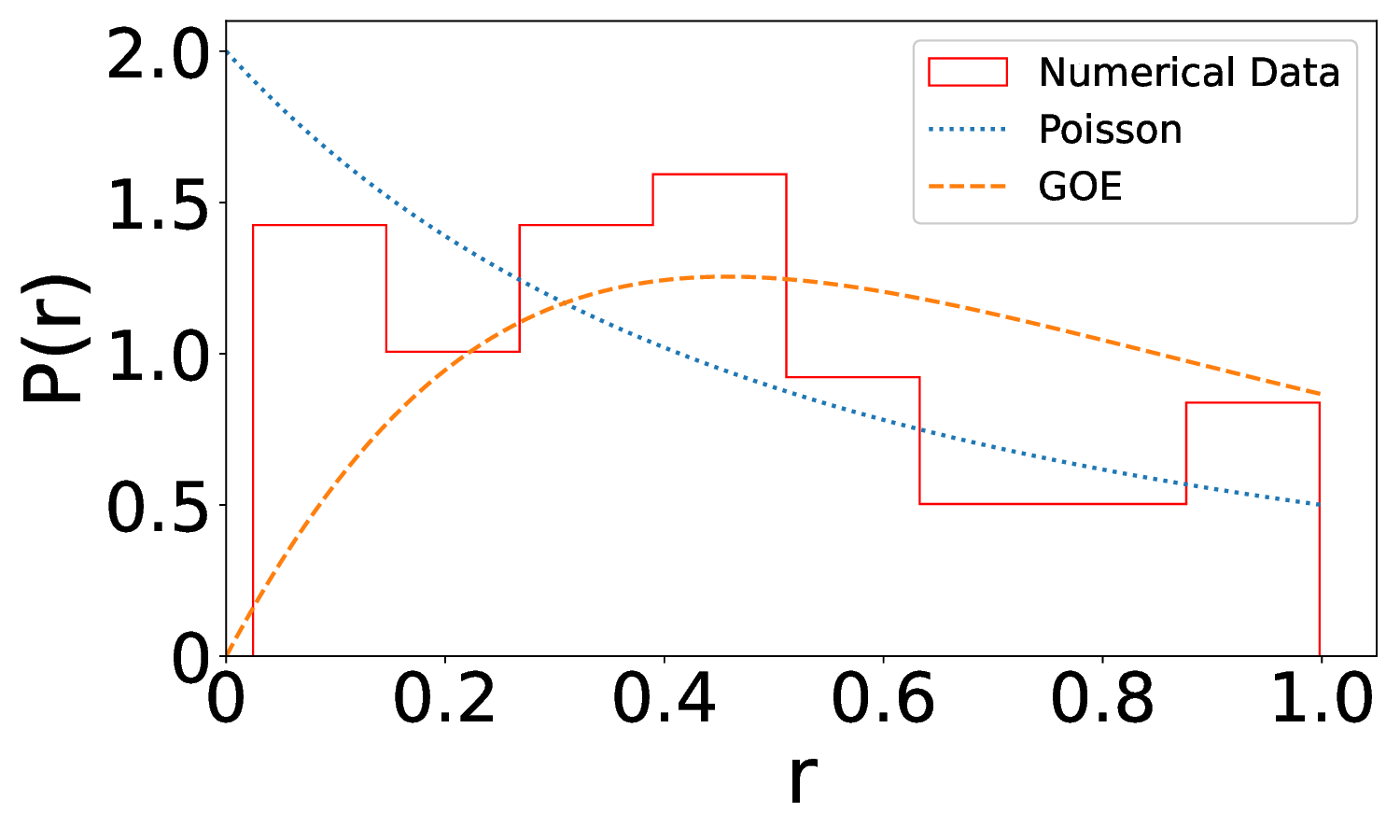}}
    
    \caption{(Color online) The nearest-neighbor spacing distribution $P(s)$ (upper panel) and the distribution of the ratio of consecutive level spacings $P(r)$ (lower panel) in the strong interaction regime with $g_{2}=3.669$ for $N=12, 16$ and $20$ bosons with total angular momentum $L_{z}=0$. The histogram in each graphs represents our numerical result for the lowest 100 energy levels. The blue dotted curve corresponds to the Poisson distribution, the orange dashed curve to the GOE distribution and the green dash-dotted curve to the Brody distribution with fitting parameter $b$.}
    \label{fig:gapnonrot}
\end{figure*}

\begin{figure*}[t]
\centering 
      \subfigure[$N=12,16$ and $20$, $g_{2}=0.3669$]{\label{fig:nonrotdeltamoderate}\includegraphics[width=0.48\linewidth]{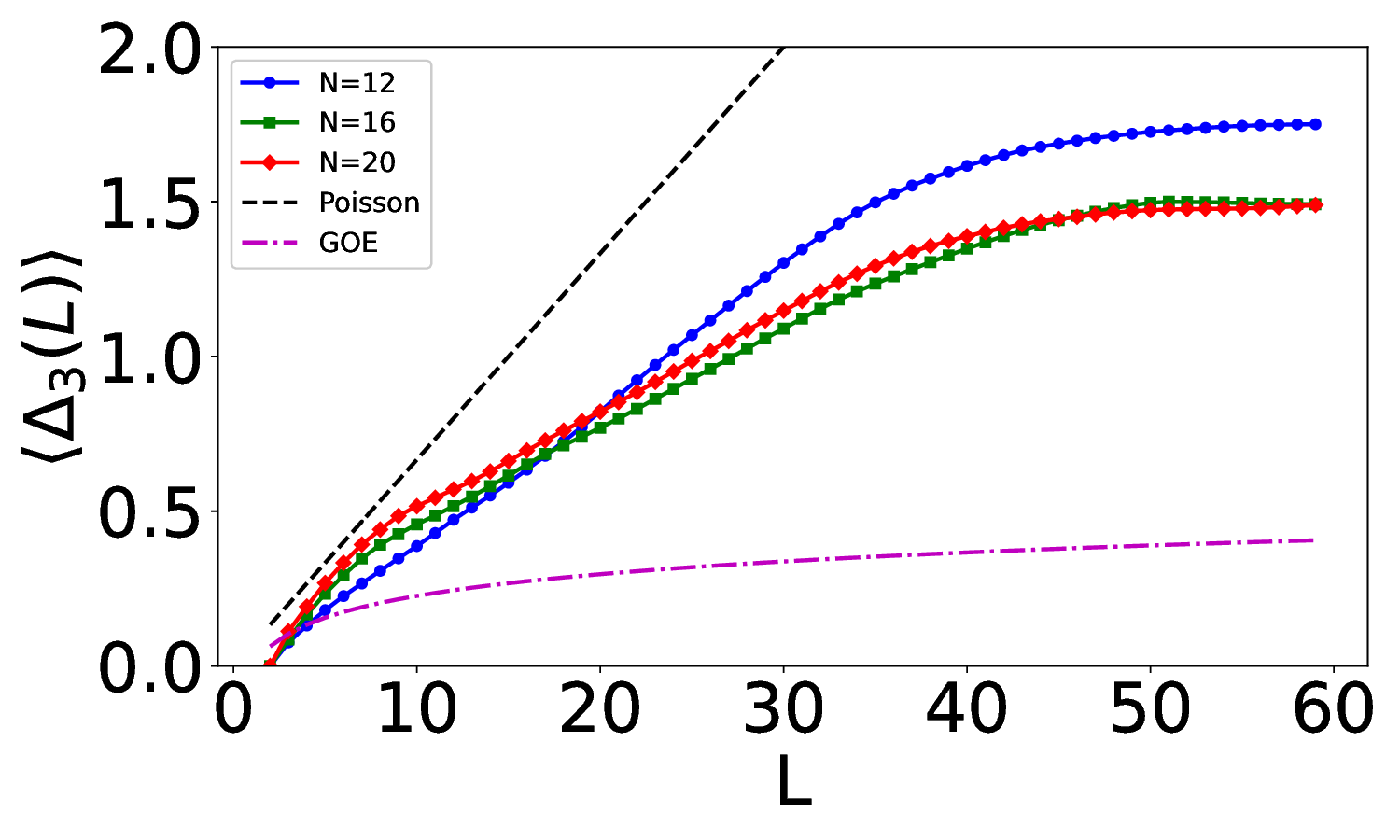}}
      \subfigure[$N=12,16$ and $20$, $g_{2}=0.3669$]{\label{fig:nonrotvariancemoderate}\includegraphics[width=0.48\linewidth]{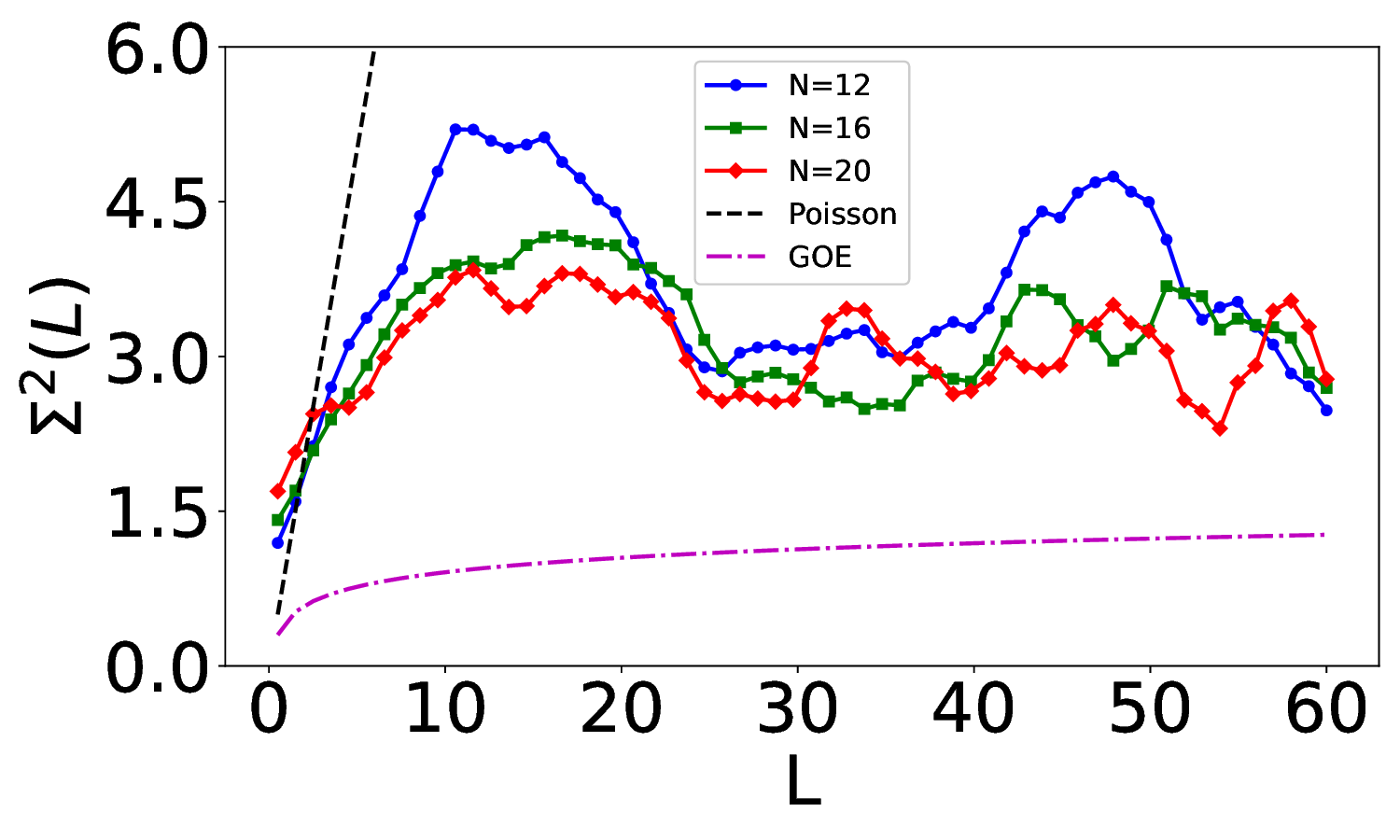}}
       \subfigure[$N=12,16$ and $20$, $g_{2}=3.669$]{\label{fig:nonrotdeltastrong}\includegraphics[width=0.48\linewidth]{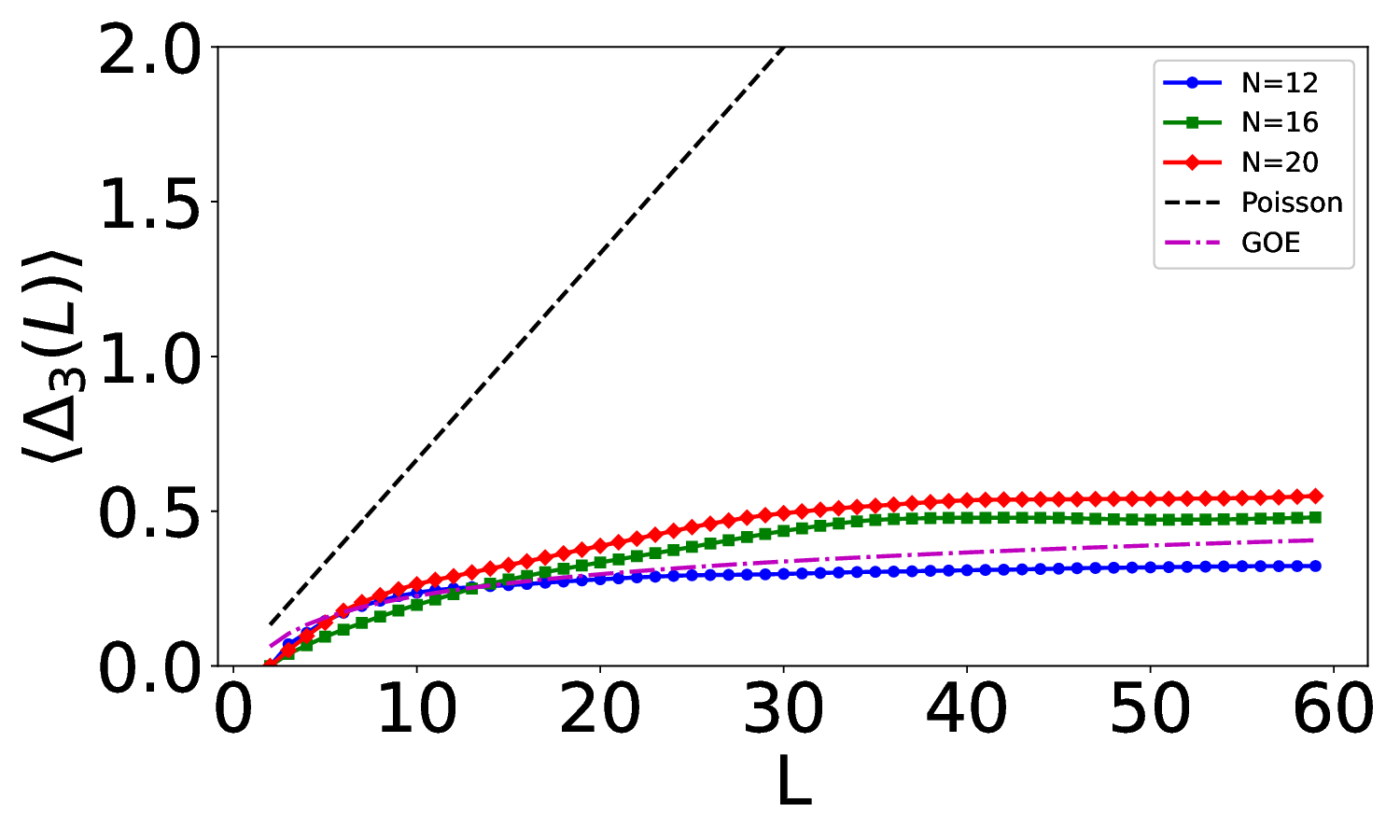}}
       \subfigure[$N=12,16$ and $20$, $g_{2}=3.669$]{\label{fig:nonrotvariancestrong}\includegraphics[width=0.48\linewidth]{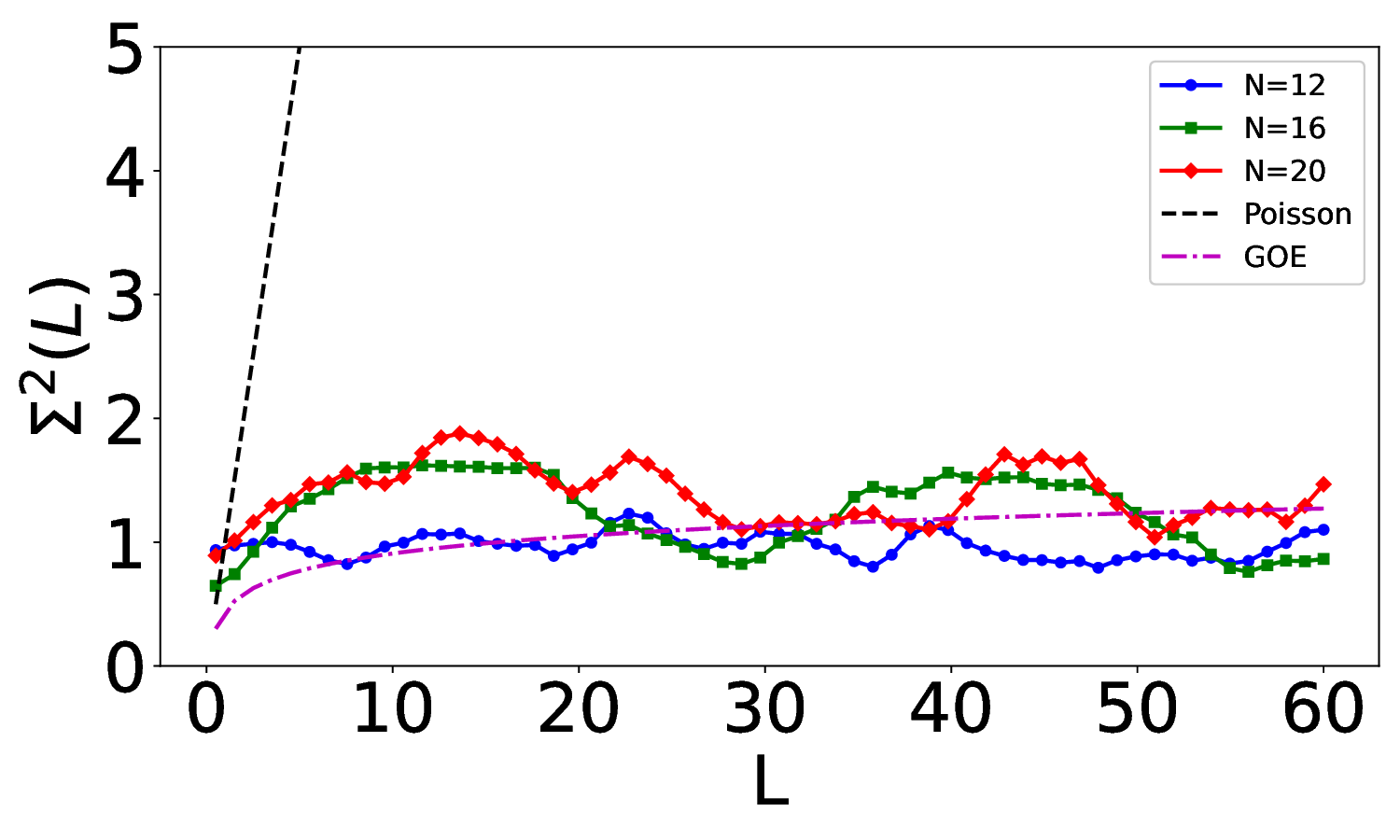}} 

    \caption{(Color online) The spectral average $\langle \Delta_3(L) \rangle$ {\it vs} $L$ and the level number variance ${\Sigma^2(L)}$ {\it vs} $L$ are presented for moderate interaction regime with $g_{2}=0.3669$ (upper panel) and strong interaction regime with $g_{2}=3.669$ (lower panel) for different number of bosons $N=12, 16$ and $20$ with total angular momentum $L_{z}=0$. The blue circle, green square and red diamond lines are our numerical results of $\langle \Delta_3(L) \rangle$ (${\Sigma^2(L)}$) for $N=12,16$ and $20$, respectively, for the lowest 100 energy levels . For reference, we have also drawn $\langle \Delta_3(L) \rangle$ (${\Sigma^2(L)}$), as a function of $L$, corressponding to the Poisson distribution (black dashed line) and the GOE distribution (magenta dash-dot line).}
    \label{fig:Deltanonrot}
\end{figure*}

\subsubsection{\bf Short-range correlations}

The NNSD $P(s)$ and the gap ratio distribution $P(r)$ provide insights into the short-range correlations. These are shown in Fig. \ref{fig:NNSDnonrotmoderate} for moderate interaction regime and in Fig. \ref{fig:gapnonrot} for the strong interaction regime, wherein the upper panel refers to the NNSD, while the lower panel corresponds to the $P(r)$ distribution. It is observed that the NNSD $P(s)$ and the $P(r)$ distribution exhibit similar behavior depending on the two-body interaction strength $g_{2}$ and the number of bosons $N$ in the system. Both the moderate and the strong interaction regimes are discussed in the following.\\

\subsubsubsection{Moderate interaction regime}

As the number of bosons increases from $N=12, 16$ to $20$, the NNSD aligns more towards  Poisson distribution as shown in Figs. \ref{fig:12brody}, \ref{fig:16brody}, and \ref{fig:20brody}, respectively (the upper panel). A more quantitative measure of the behavior is given by the Brody parameter $b$ obtained by fitting Eq. (\ref{fittingbrody}). In Table \ref{tablebrody1}, we present the Brody parameter $b$ for $N=12, 16$ and $20$, which in the moderate interaction regime (with $g_{2}=0.3669$) is found to decrease systematically with increase in number of bosons (third subcolumn). For $N=12$, the Brody parameter is $b=0.33$, signifying small correlations between the energy levels. However, as the number of bosons increases to $N=16$, the Brody parameter decreases to $b=0.16$ and for $N=20$, the Brody parameter further decreases to $b=0.04$, indicating a Poissonian distribution where the level repulsion is minimum. The observed behavior implies that an increase in number of bosons $N$ leads to decrease in spectral correlations. This shows that, in this regime, the energy levels remain largely uncorrelated, a characteristic feature of a regular system.

\begin{table}[t]
\caption{The mean gap ratio $\langle r \rangle$ with estimated error for different number of bosons in the non-rotating case in the moderate ($g_{2}=0.3669$) and the strong ($g_{2}=3.669$) interaction regimes.}
    \label{tablegapratio1}
	\centering
	\begin{tabular}{|c|c|c|c|} \hline
		& Number of bosons, $N$ & \multicolumn{2}{c|}{Mean gap ratio $\langle r \rangle$} \\ \cline{3-4}
		& & $g_{2}=0.3669$ & $g_{2}=3.669$ \\ \hline
		& 12 & 0.398 $\pm$ 0.027 & 0.442 $\pm$ 0.028 \\ \hline
		& 16 & 0.403 $\pm$ 0.030 & 0.538 $\pm$ 0.028 \\ \hline
		& 20 & 0.374 $\pm$ 0.028 & 0.435 $\pm$ 0.027 \\ \hline
		Poisson &  & \multicolumn{2}{c|}{0.386} \\ \hline
		GOE     &  & \multicolumn{2}{c|}{0.530} \\ \hline
	\end{tabular}
\end{table}

\begin{table}[t]
	\caption{Fitted values of the Brody parameter $b$ in the rotating single-vortex state $L_{z}=N$ for different number of bosons in the moderate ($g_{2}=0.3669$) and the strong ($g_{2}=3.669$) interaction regimes.}
    \label{tablebrodyrot2}
	\centering
	\begin{tabular}{|c|c|c|c|} \hline
		& Number of bosons, $N$ & \multicolumn{2}{c|}{Brody parameter $b$} \\ \cline{3-4}
		& & \textbf{$g_2 = 0.3669$} & \textbf{$g_2 = 3.669$} \\ \hline
		& 12 & 0.52 & 1.18 \\ \hline
		& 16 & 0.64 & 0.98 \\ \hline
		& 20 & 0.64 & 1.07 \\ \hline
		Poisson &    & \multicolumn{2}{c|}{0} \\ \hline
		GOE     &    & \multicolumn{2}{c|}{1} \\ \hline
	\end{tabular}
\end{table}
 
The results for the consecutive ratios of level spacings distribution $P(r)$ with increasing number of bosons $N = 12$, $16$ and $20$ are presented in Figs. \ref{fig:12gap}, \ref{fig:16gap} and \ref{fig:20gap}, respectively (the lower panel). The corresponding values of the mean gap ratio $\langle r \rangle$ are summarized in the third subcolumn of Table \ref{tablegapratio1}. For $N=12$, the calculated mean gap ratio is $\langle r \rangle=0.398$. As the number of bosons increases to $N=16$, the mean gap ratio shows a marginal increase to $\langle r \rangle=0.403$. Upon further increase in the number of bosons to $N=20$, the mean gap ratio becomes $\langle r \rangle=0.374$. These values of $\langle r \rangle$ are in close agreement with the theoretical value of $\langle r \rangle\approx0.386$ for a Poissonian distribution, indicating that the system primarily features regular behavior with no significant spectral correlations. The mean gap ratio for the two subsets of lowest energy levels—the lower 1-50 and the higher 51–100 levels—is shown in fourth column of Table \ref{gapratio50-100}. It is observed that the lower levels ($1$–$50$) exhibit regular behavior, as the mean gap ratio is close to the theoretical Poisson value of $0.386$. The higher levels ($51$–$100$) also show regular behavior with a slightly larger mean gap ratio compared to the lower levels.

The observations from the NNSD and the $P(r)$ distribution suggest that, in the moderate interaction regime, the level statistics is well described by a Poisson distribution. This behavior signifies the absence of level repulsion, leading to the conclusion that the system exhibits regular, non-chaotic behavior.\\

\begin{table}[t]
	\caption{The mean gap ratio $\langle r \rangle$ with estimated error for different number of bosons in the rotating single-vortex state $L_{z}=N$ in the moderate ($g_{2}=0.3669$) and the strong ($g_{2}=3.669$) interaction regimes.}
    \label{tablegapratiorot2}
	\centering
	\begin{tabular}{|c|c|c|c|} \hline
		& Number of bosons, $N$ & \multicolumn{2}{c|}{Mean gap ratio $\langle r \rangle$} \\ \cline{3-4}
		& & \textbf{$g_2 = 0.3669$} & \textbf{$g_2 = 3.669$} \\ \hline
		& 12 & 0.498 $\pm$ 0.027 & 0.555 $\pm$ 0.026 \\ \hline
		& 16 & 0.494 $\pm$ 0.028 & 0.505 $\pm$ 0.027 \\ \hline
		& 20 & 0.496 $\pm$ 0.025 & 0.549 $\pm$ 0.023 \\ \hline
		Poisson &    & \multicolumn{2}{c|}{0.386} \\ \hline
		GOE     &    & \multicolumn{2}{c|}{0.530} \\ \hline
	\end{tabular}
\end{table}

\subsubsubsection{Strong interaction regime}

The results on NNSD $P(s)$ for varying numbers of bosons $N = 12, 16, 20$ are presented in Figs. \ref{fig:12brody9.151}–\ref{fig:20brody9.151}, in the strong interaction regime ($g_{2} = 3.669$). For $N = 12$, the NNSD follows the GOE distribution with Brody parameter $b = 0.53$, indicating small spectral correlations leading to a weakly chaotic behavior, as shown in Fig. \ref{fig:12brody9.151}. This suggests that while signatures of chaotic behavior are present, a degree of regularity persists within the system. As the number of bosons increases to $N = 16$, the spectral correlation between energy levels enhances, as evidenced by increase in Brody parameter to the value $b = 0.85$, shown in Fig. \ref{fig:16brody9.151}. This considerable increase in Brody parameter implies that the system has developed a significantly chaotic behavior, exhibiting strong level repulsion, a characteristic of GOE statistics. However, for $N = 20$, the Brody parameter decreases to $b = 0.51$, as seen in Fig. \ref{fig:20brody9.151}, indicating that the system returns to a weakly chaotic regime with revival of some degree of regularity within the system. The corresponding values of the Brody parameter $b$ are summarized in the fourth subcolumn of Table \ref{tablebrody1}.

\begin{table*}[t]
\caption{The mean gap ratio with estimated error for the two subsets of energy levels—the lower 1–50 and the higher 51–100 levels—is calculated for different number of bosons in the non-rotating case ($L_{z}=0$) in both moderate ($g_{2}=0.3669$) and strong ($g_{2}=3.669$) interaction regimes.}
\label{gapratio50-100}
\centering
\begin{tabular}{|c|c|c|c|c|} 
\hline
 & Number of bosons,  $N$ & Number of levels & \multicolumn{2}{c|}{Mean gap ratio $\langle r \rangle$} \\ 
\cline{4-5}
 & & & \textbf{$g_2 = 0.3669$} & \textbf{$g_2 = 3.669$} \\ 
\hline
 & \multirow{2}{*}{12} & 1--50 & 0.363 $\pm$ 0.040  & 0.411 $\pm$ 0.042 \\ 
\cline{3-5}
 &  & 51--100 & 0.432 $\pm$ 0.037 & 0.472 $\pm$ 0.038 \\ 
\hline
 & \multirow{2}{*}{16} & 1--50 & 0.331 $\pm$ 0.039 & 0.572 $\pm$ 0.043 \\ 
\cline{3-5}
 &  & 51--100 & 0.478 $\pm$ 0.044 & 0.499 $\pm$ 0.035 \\ 
\hline
 & \multirow{2}{*}{20} & 1--50 & 0.338 $\pm$ 0.038 & 0.492 $\pm$ 0.041 \\ 
\cline{3-5}
 &  & 51--100 & 0.423 $\pm$ 0.042 & 0.382 $\pm$ 0.035 \\ 
\hline
Poisson &  &  & \multicolumn{2}{c|}{0.386} \\ 
\hline
GOE     &  &  & \multicolumn{2}{c|}{0.530} \\ 
\hline
\end{tabular}
\end{table*}

The distribution of the ratio of consecutive level spacings $P(r)$ for the increasing number of bosons $N=12$, $16$ to $20$ is shown in Figs. \ref{fig:12gap9.151}–\ref{fig:20gap9.151}. For $N=12$,  the $P(r)$ distribution begins to deviate from Poissonian statistics and shows characteristics similar to those of GOE, as seen in Fig. \ref{fig:12gap9.151}. The corresponding mean gap ratio is $\langle r \rangle=0.442$, indicating an intermediate regime in which the system exhibits signatures of weak chaotic behavior with some degree of regularity. As the number of bosons increases to $N=16$, the $P(r)$ distribution aligns closely with GOE distribution, as seen in Fig. \ref{fig:16gap9.151}. The calculated mean gap ratio in this case is $\langle r \rangle=0.538$, which is close to the theoretical value of $\langle r \rangle \approx 0.530$ for GOE distribution. This indicates the emergence of strong spectral correlations, suggesting significant quantum-chaotic behavior in the system. As the number of bosons increases further to $N=20$, the $P(r)$ distribution returns to align with GOE distribution with revival of some degree of regularity in the system, as shown in Fig. \ref{fig:20gap9.151}. The calculated mean gap ratio decreases to $\langle r \rangle=0.435$, implying a weakening of chaotic behavior. The values of mean gap ratio $\langle r \rangle$ corresponding to each number of bosons $N$ are summarized in the fourth subcolumn of Table \ref{tablegapratio1}. The mean gap ratio for the lower ($1$–$50$) and higher ($51$–$100$) subsets of energy levels is summarized in the fifth column of Table \ref{gapratio50-100}. The lower ($1$–$50$) part of the spectrum exhibits chaotic behavior consistent with GOE distribution (mean gap ratio, $\langle r \rangle \approx 0.530$). The higher ($51$–$100$) levels retain signatures of chaoticity, except for $N=20$, where the mean gap ratio shows deviation from GOE distribution to align with Poisson distribution.

Thus, in the strongly interacting regime, the system goes over to chaotic regime with the degree of chaos modulated by the number of bosons in the system. Additionally, the interplay between the interaction strength and the number of bosons also play a significant role.

\subsubsection{\bf Long-range correlations}
To investigate the long-range correlations in the system, we employ the Dyson-Mehta $\Delta_3(L)$ statistic and the level number variance $\Sigma^2(L)$, in the moderate as well as the strong interaction regimes. These statistical measures give important insight into the spectral rigidity present in the quantum many-body systems.\\

\subsubsubsection{Moderate interaction regime}

The spectral average $\Delta_3(L)$ for moderate interaction regime is calculated over the interval $L=2$ to $60$, where $L$ denotes the length of the energy interval. We linearly fit our numerical data for $\Delta_3(L)$ to obtain the slope of $\Delta_3(L)$ {\it vs} $L$, upto $L=10$ and compare it with the theoretical value $\Delta_3(L)=\frac{L}{15}=0.0667L$ of Poisson distribution. 

For $N=12$, the $\Delta_3(L)$ statistic exhibits a linear increase with a slope of 0.0467 up to $L=10$, beyond which it saturates to a constant value, as shown in Fig. \ref{fig:nonrotdeltamoderate}. A similar trend is observed for $N=16$ and $20$, where the corresponding slopes are 0.0569 and 0.0634, respectively, as shown in Fig. \ref{fig:nonrotdeltamoderate}. Thus, as the number of bosons increases, the behavior of $\Delta_3(L)$ agrees more with Poisson distribution upto energy interval $L=10$, indicating decrease in spectral correlations in the system. At large values of $L$, the $\Delta_3(L)$ statistic saturates to $\Delta_{\infty}=1.74$, $1.50$, and $1.48$ for $N=12$, $16$, and $20$, respectively, as shown in Fig.~\ref{fig:nonrotdeltamoderate}. This behavior is consistent with Berry’s argument~\cite{berryrigidity} that $\Delta_3(L)$ approaches a constant value in the large $L$ limit.

Another indicator of long-range spectral correlation is the level number variance $\Sigma^2(L)$. We perform a linear fit to our numerical data to determine the slope of $\Sigma^2(L)$ {\it vs} $L$, up to $L=2$ and compare it with the theoretical value $\Sigma^2(L)=L$ of Poisson distribution. For $N = 12, 16$ and $20$, the corresponding slope values are $0.6615, 0.7561$ and $0.9402$, respectively, indicating an increasing alignment with Poisson distribution, which has slope of $1$, as shown in Fig. \ref{fig:nonrotvariancemoderate}. The increasing slope with increasing number of bosons $N$ suggests a reduction in spectral correlations in the system. Thus, $\Sigma^2(L)$, on the average, closely follows Poisson distribution, up to $L=2$, while for $L >2$, it exhibits noticeable deviation, as shown in Fig. \ref{fig:nonrotvariancemoderate}.\\

\subsubsubsection{Strong interaction regime}

The spectral average $\Delta_3(L)$ for different number of bosons $N$, in strong interaction regime, is shown in Fig. \ref{fig:nonrotdeltastrong}. Our numerical result demonstrates that the $\Delta_3(L)$ statistic follows the GOE distribution up to a certain range of the energy interval $L$, beyond which it saturates to a constant value $\Delta_{\infty}=0.321, 0.475$ and $0.543$ for $N=12, 16$ and $20$, respectively. Notably, this saturation for the strong interaction occurs at a significantly lower value compared to the moderate interaction case, where $\Delta_{\infty}=1.74, 1.50$ and $1.48$ for the respective number of bosons. This is consistent with Berry's argument that integrable systems exhibit higher saturated $\Delta_3$ values than non-integrable (chaotic) ones \cite{berryrigidity}. For example, in the non-rotating case with $N=12$, the saturated values of $\Delta_3$ are found to be 1.74 and 0.321 for the moderate and strong interaction regimes, respectively. This suggests that the system is integrable in the moderate interaction regime and becomes non-integrable in the strong interaction regime. Similar trend is observed for $N=16$ and $N=20$ across both interaction regimes. As the number of bosons increases from $N=12, 16$, $20$, the saturation value of $\Delta_3(L)$ at large $L$ also increases, as shown in Fig. \ref{fig:nonrotdeltastrong}. 

\vspace{-1pt}The variation of the level number variance $\Sigma^2(L)$ with increasing number of bosons is depicted in Fig. \ref{fig:nonrotvariancestrong}. For small $L$, $\Sigma^2(L)$ closely follows the GOE distribution. However, at larger $L$, the deviation of $\Sigma^2(L)$ from the GOE distribution becomes more pronounced as the number of bosons increases.

\begin{figure*}[t]
    \centering  
         \subfigure[$N=12$, $L_{z}=N$]{\label{fig:12brodyrotatingcase}\includegraphics[width=0.32\linewidth]{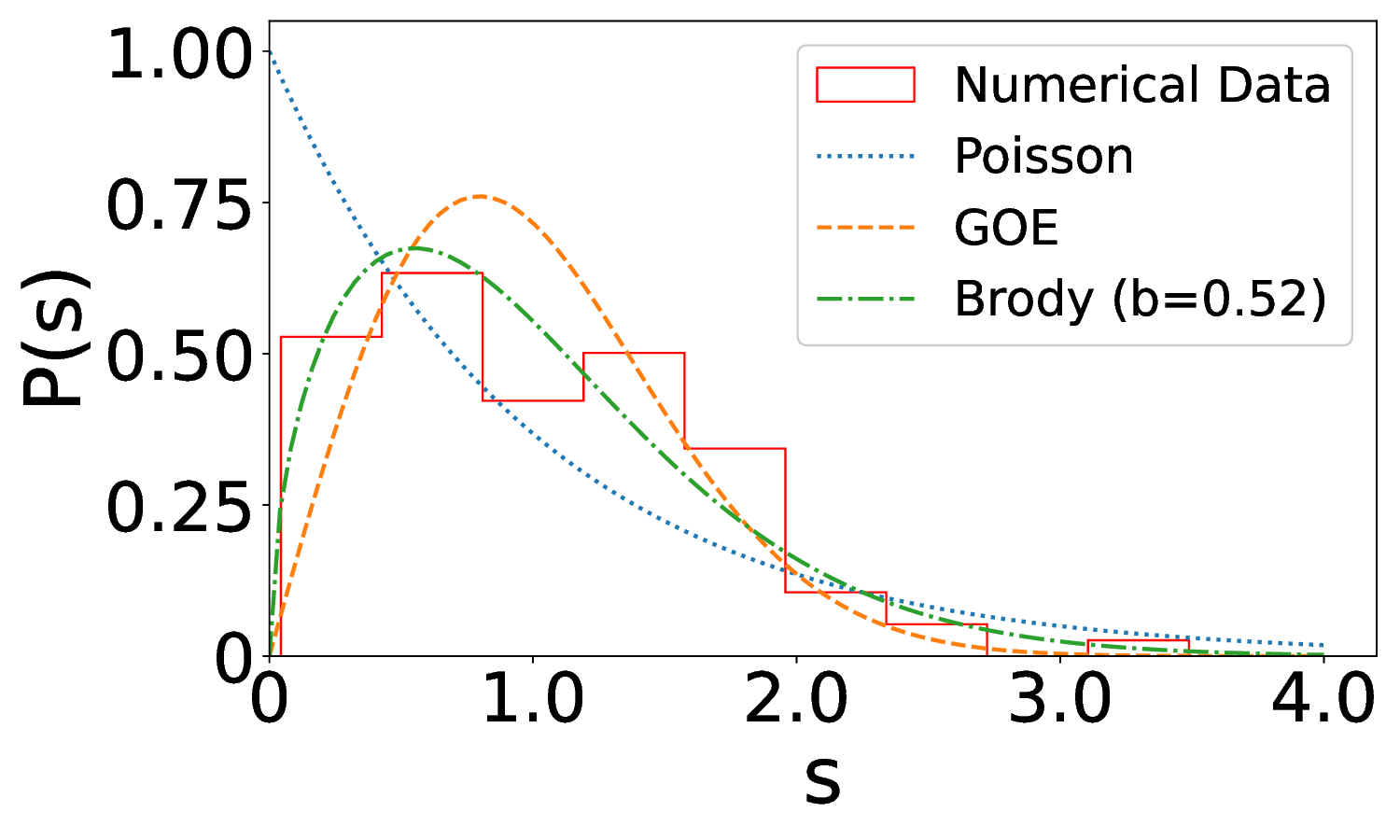}}
        \subfigure[$N=16$, $L_{z}=N$]{\label{fig:24brodyrotatingcase}\includegraphics[width=0.32\linewidth]{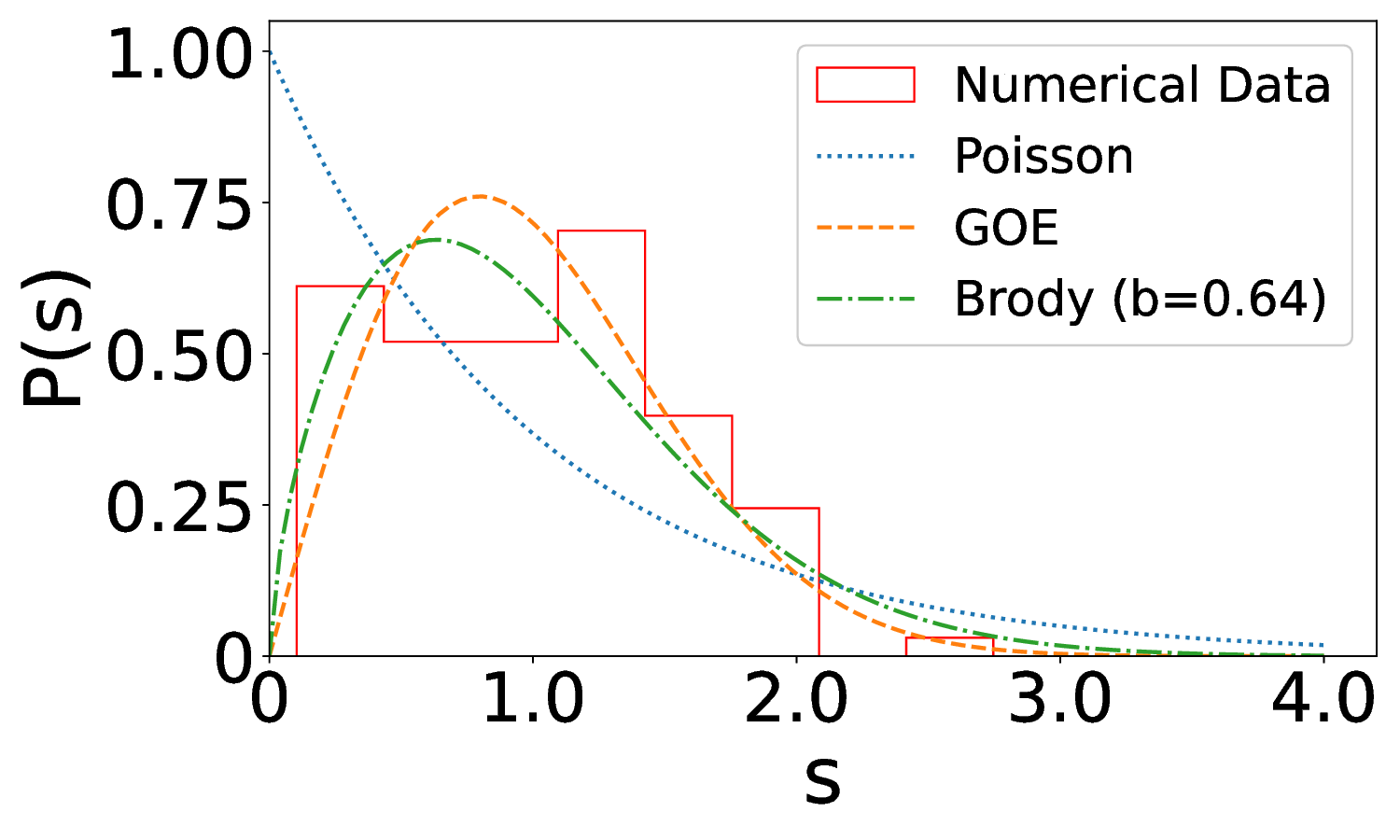}}
        \subfigure[$N=20$, $L_{z}=N$]{\label{fig:36brodyrotatingcase}\includegraphics[width=0.32\linewidth]{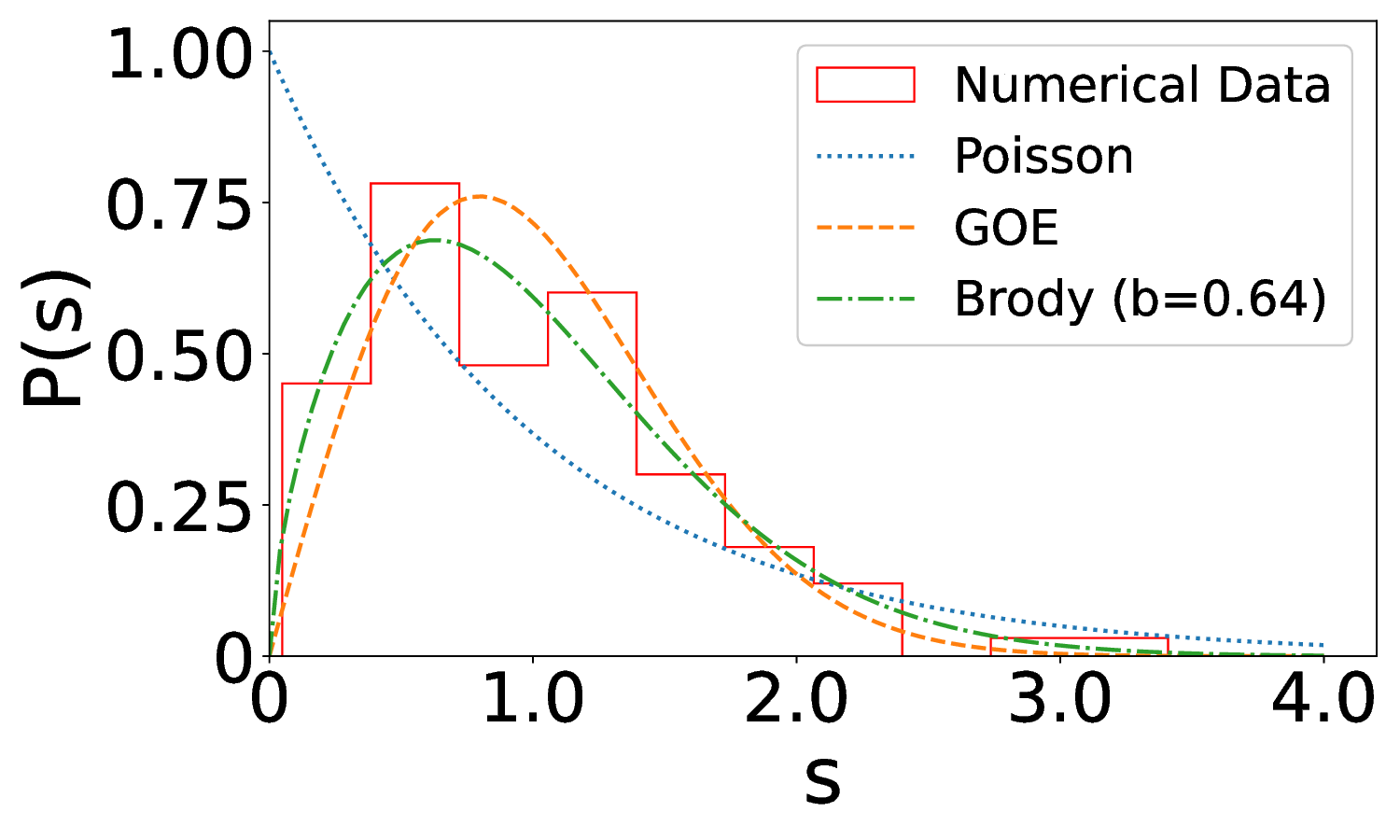}}
        \subfigure[$N=12$, $L_{z}=N$]{\label{fig:12gaprotatingcase}\includegraphics[width=0.32\linewidth]{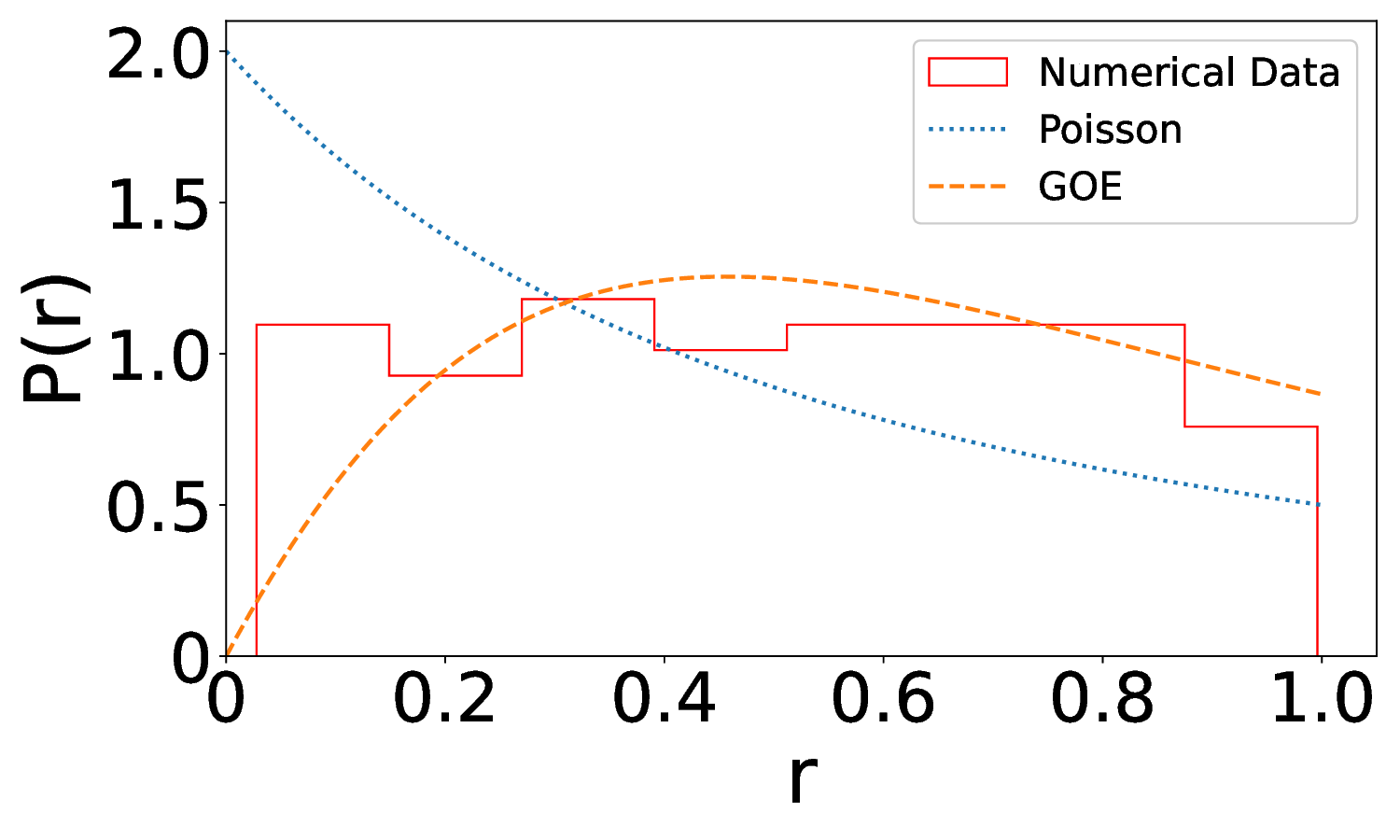}}
        \subfigure[$N=16$, $L_{z}=N$]{\label{fig:24gaprotatingcase}\includegraphics[width=0.32\linewidth]{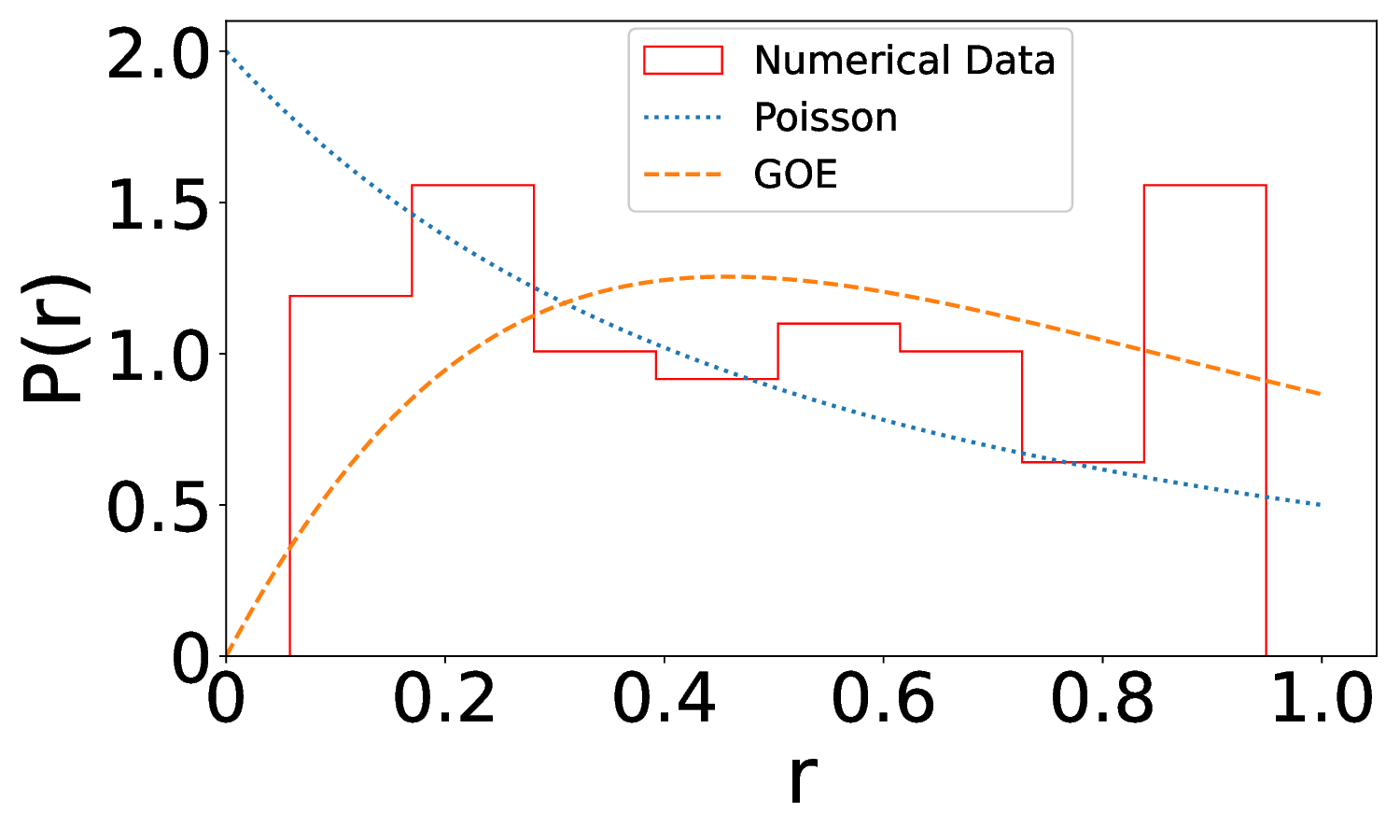}}
        \subfigure[$N=20$, $L_{z}=N$]{\label{fig:36gaprotatingcase}\includegraphics[width=0.32\linewidth]{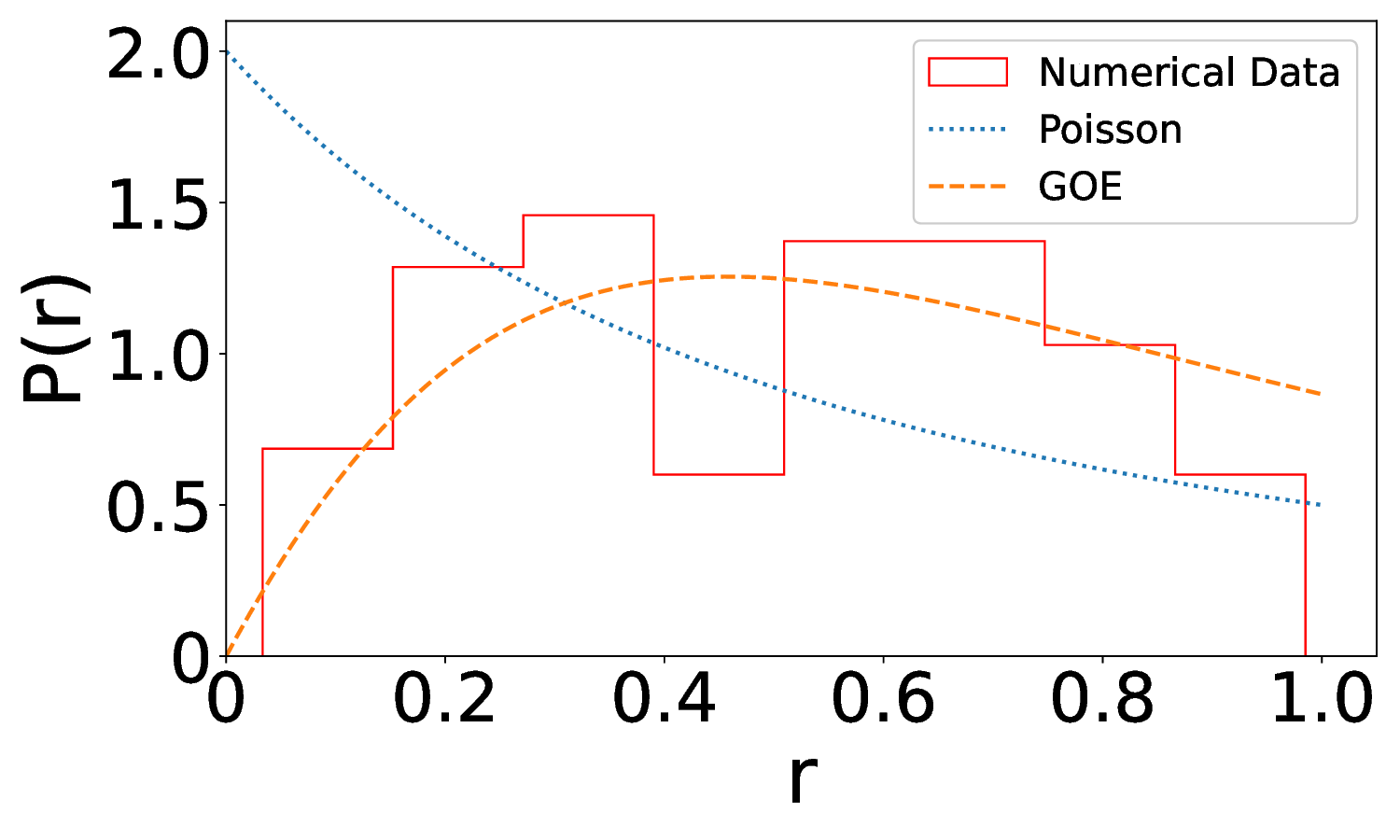}}
 \caption{(Color online) The nearest-neighbor spacing distribution $P(s)$ (upper panel) and the distribution of the ratio of consecutive level spacings $P(r)$ (lower panel) for moderate interaction regime with $g_{2}=0.3669$ in the single-vortex state $L_{z}=N$ where $N=12,16$ and $20$. The histogram in each graphs represents our numerical result for the lowest 100 energy levels. The blue line corresponds to the Poisson distribution, the orange dashed curve to the GOE distribution, and the green dash-dotted curve to the Brody distribution with fitting parameter $b$.}
    \label{fig:rotatingcasemoderate}
\end{figure*}

\subsection{\label{Rcase}Rotating case }
We focus on the rotating single-vortex state $L_{z}=N$ for number of bosons $N=12, 16$ and $20$ for both the short-range and the long-range correlations.

\begin{figure*}[t]
\centering
\subfigure[$N=12$, $L_{z}=N$]{%
    \label{fig:12nnsdrot7.301}
    \includegraphics[width=0.32\linewidth]{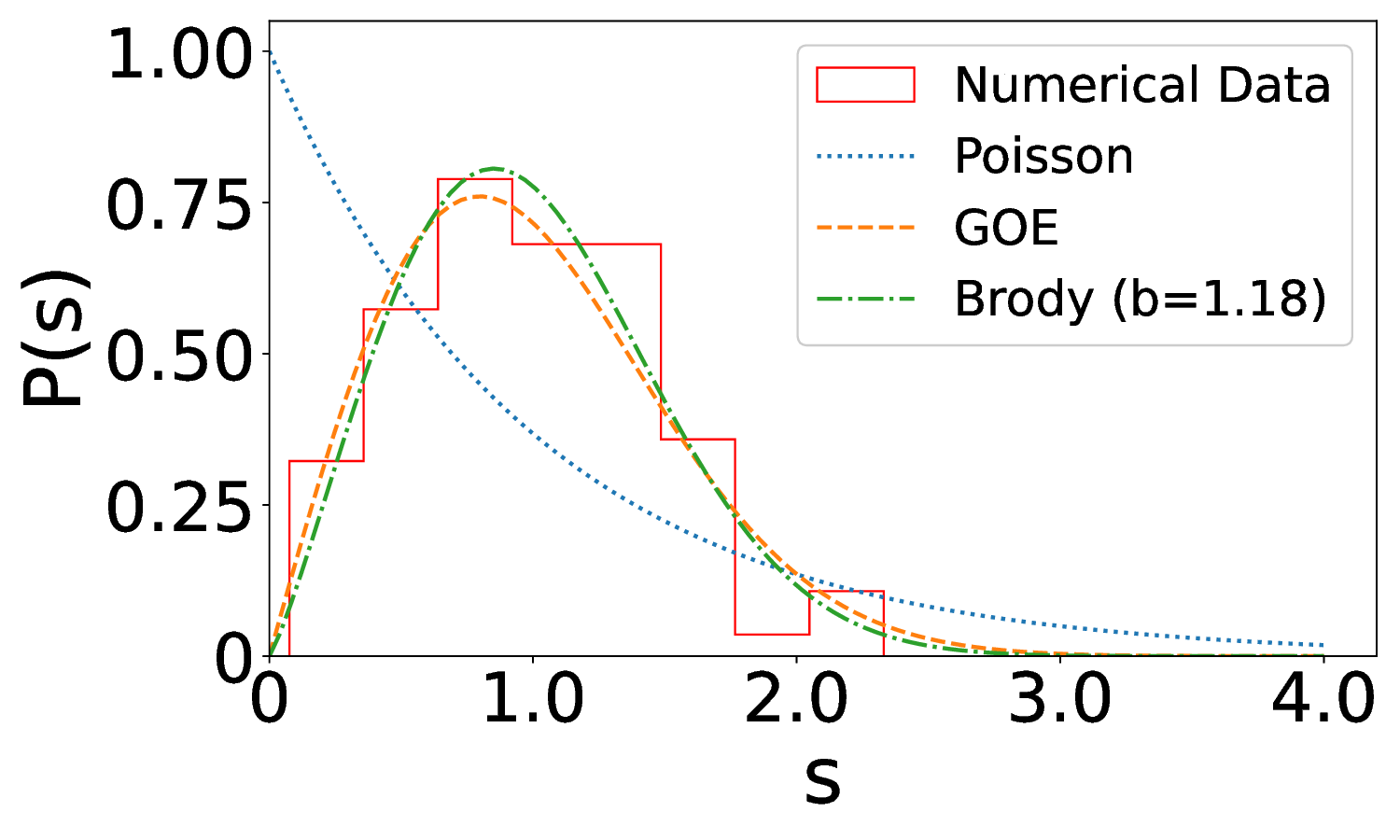}}
\subfigure[$N=16$, $L_{z}=N$]{%
    \label{fig:24nnsdrot7.301}
    \includegraphics[width=0.32\linewidth]{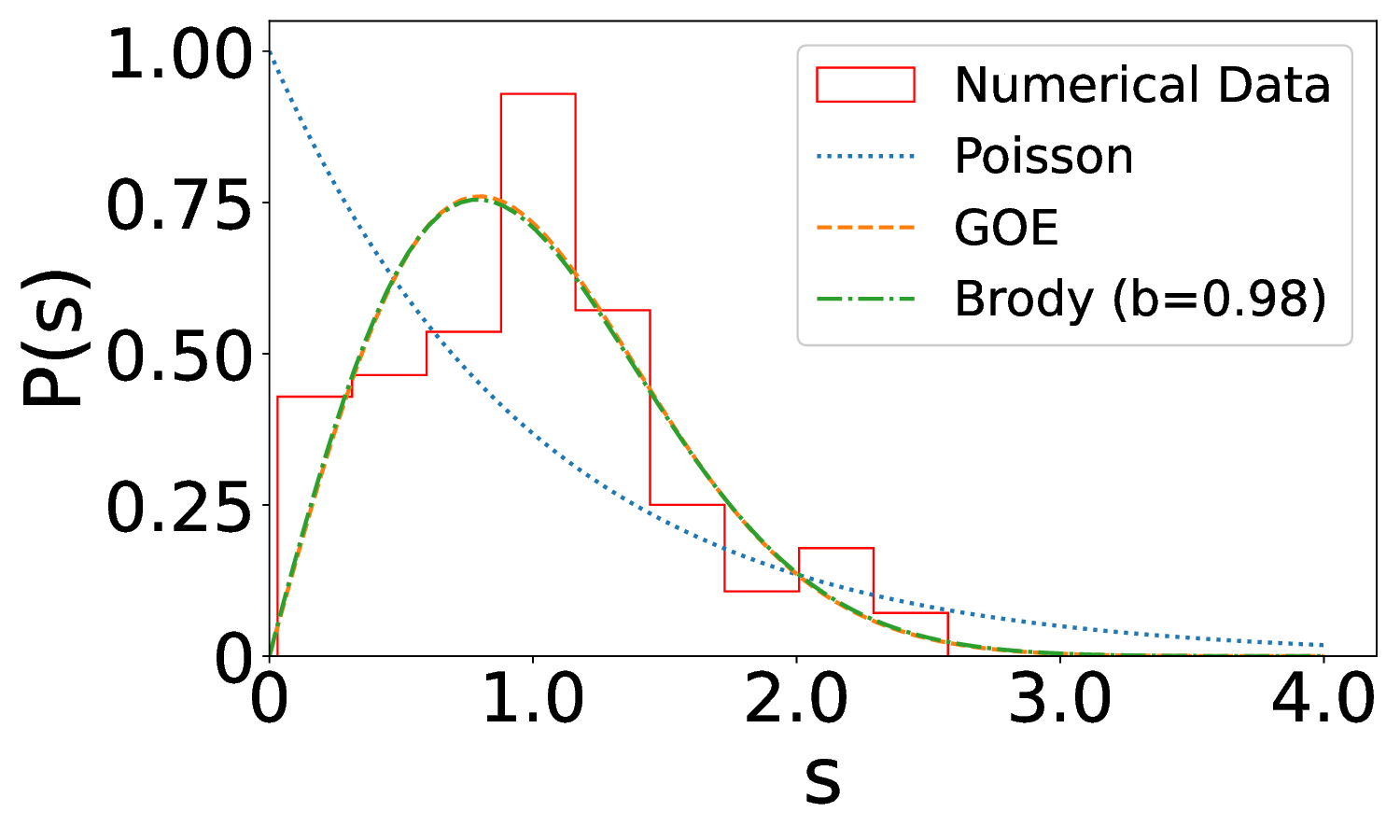}}
\subfigure[$N=20$, $L_{z}=N$]{%
    \label{fig:36nnsdrot7.301}
    \includegraphics[width=0.32\linewidth]{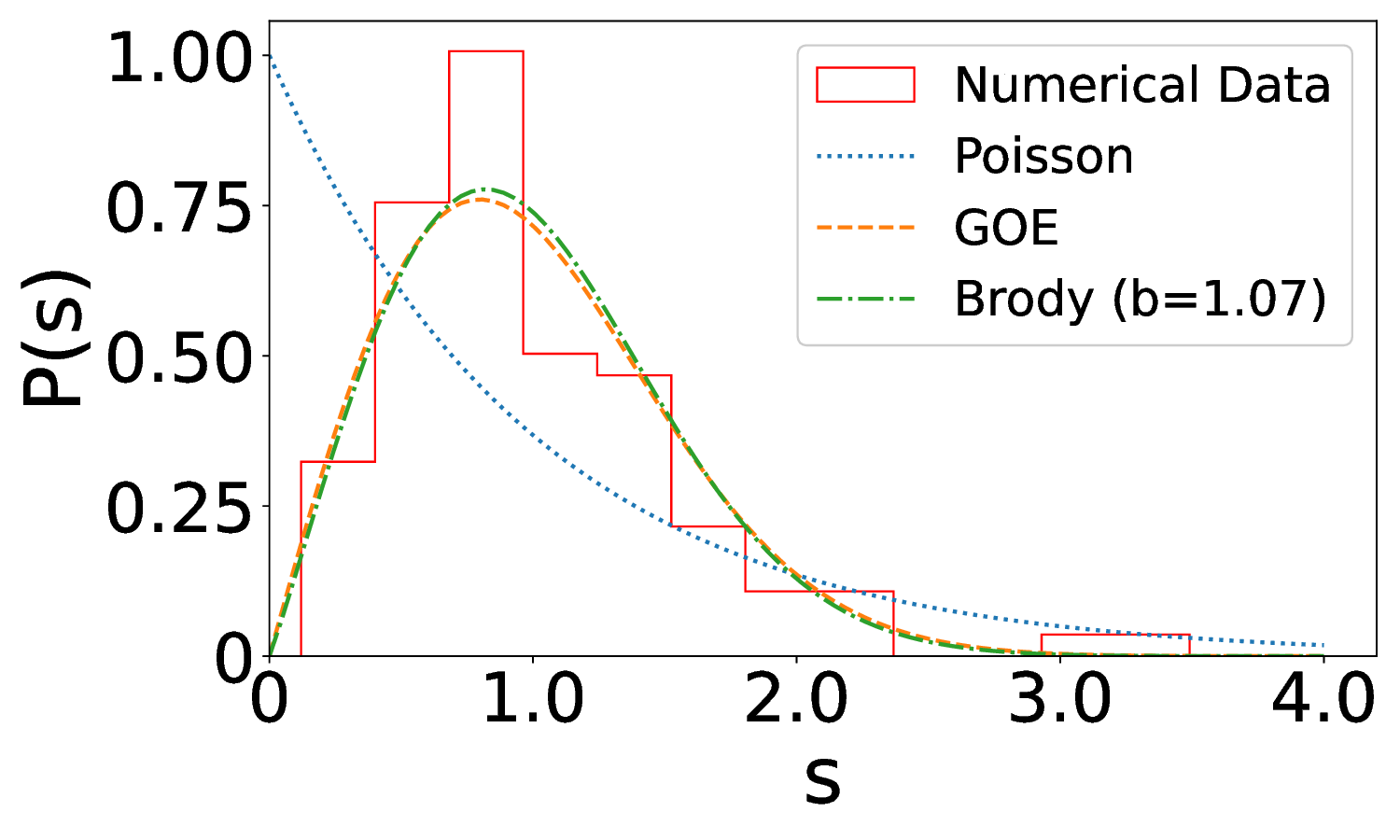}}

\subfigure[$N=12$, $L_{z}=N$]{%
    \label{fig:12gaprot7.301}
    \includegraphics[width=0.32\linewidth]{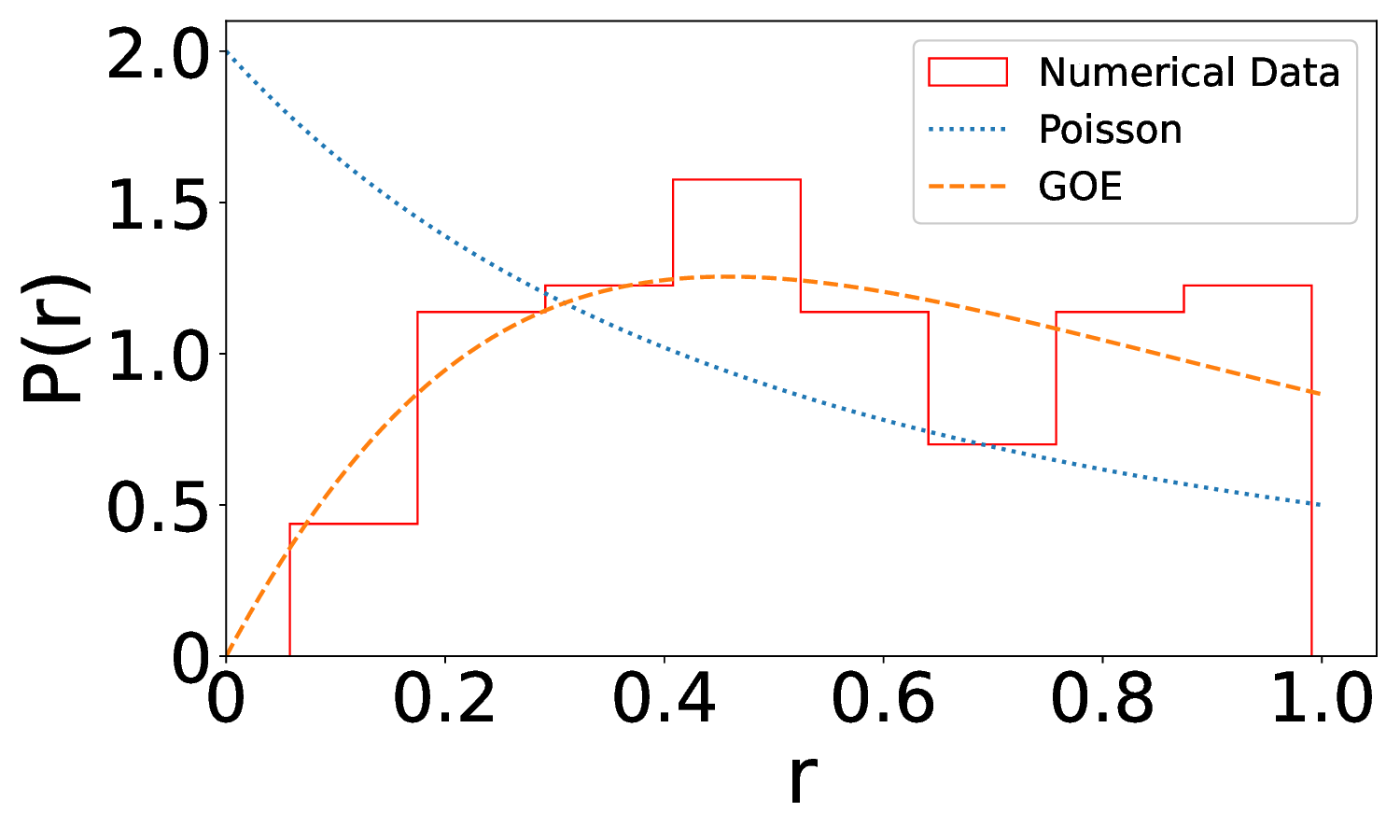}}
\subfigure[$N=16$, $L_{z}=N$]{%
    \label{fig:24gaprot7.301}
    \includegraphics[width=0.32\linewidth]{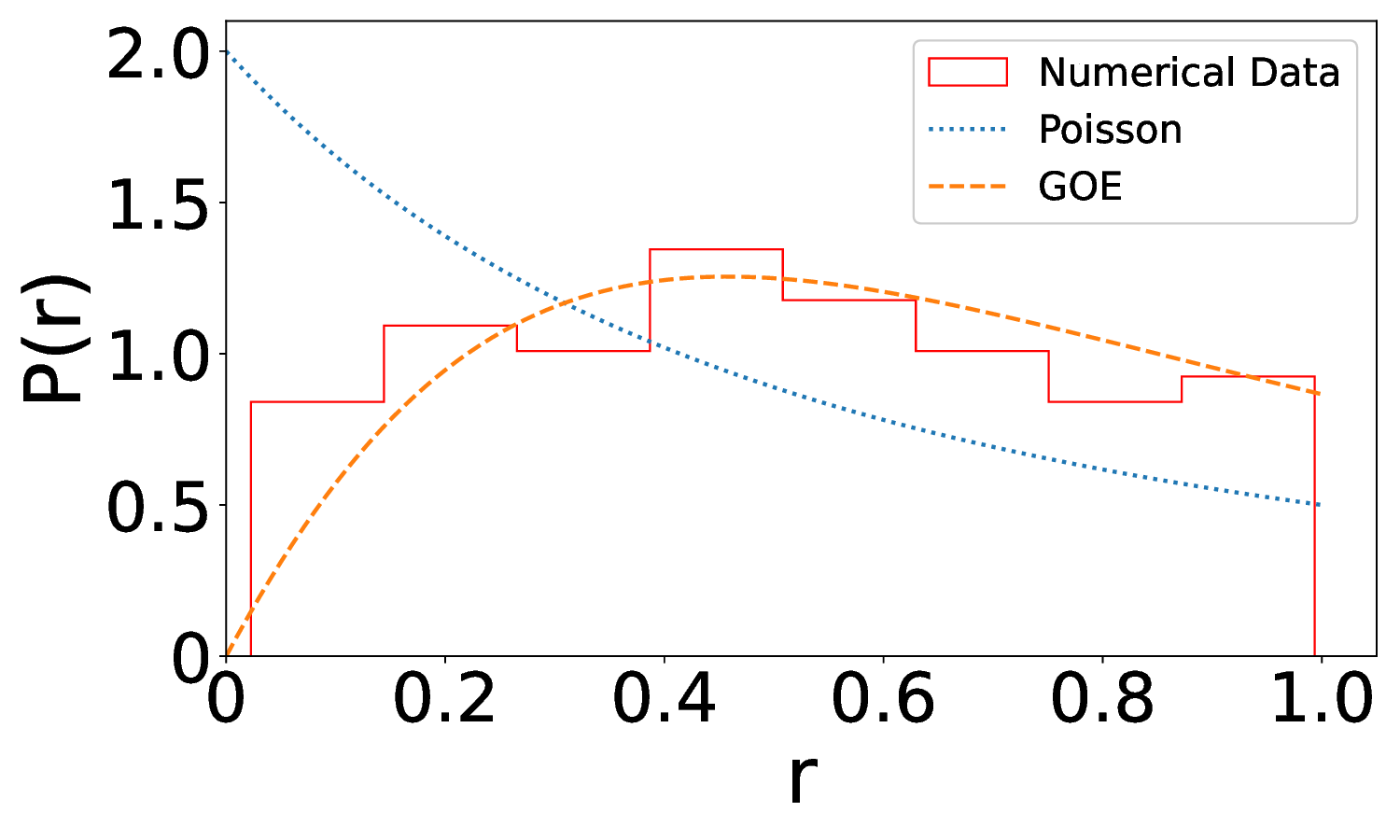}}
\subfigure[$N=20$, $L_{z}=N$]{%
    \label{fig:36gaprot7.301}
    \includegraphics[width=0.32\linewidth]{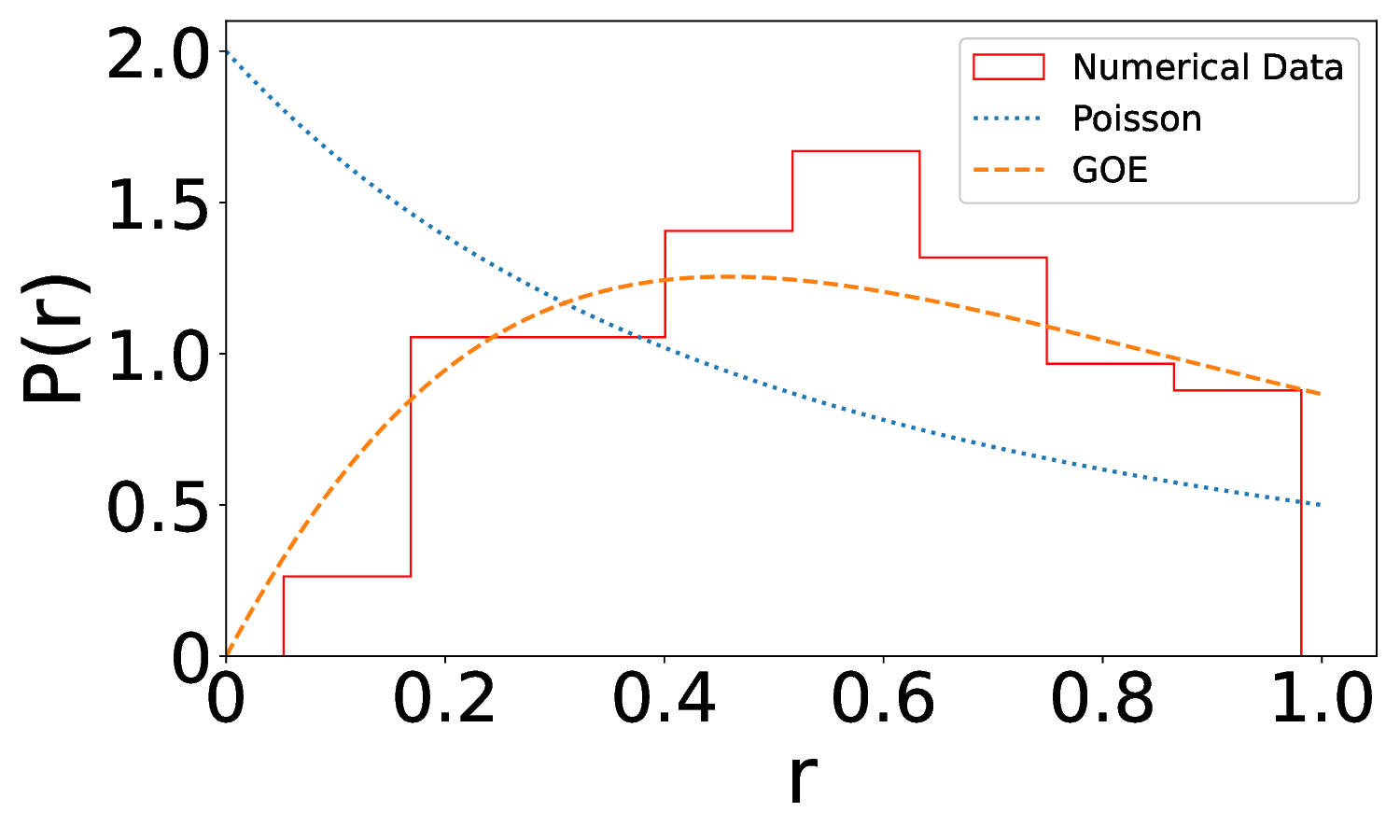}}

%\vspace{10pt}
\caption{(Color online) Upper panel: Nearest-neighbor spacing distribution $P(s)$; Lower panel: Distribution of the ratio of consecutive level spacings $P(r)$ for the strong interaction regime with $g_{2}=3.669$ in the single-vortex state $L_{z}=N$ for $N=12$, $16$, and $20$. The histogram in each graph represents our numerical result for the lowest 100 energy levels. The blue line shows Poisson, the orange dashed line GOE, and the green dash-dotted line the Brody distribution with fitted parameter $b$.}
\label{fig:gaprot7.301}
\end{figure*}

\begin{figure*}[t]
\centering
\subfigure[$L_{z}=N$, $g_{2}=0.3669$]{%
    \label{fig:rotdeltamoderate}
    \includegraphics[width=0.48\linewidth]{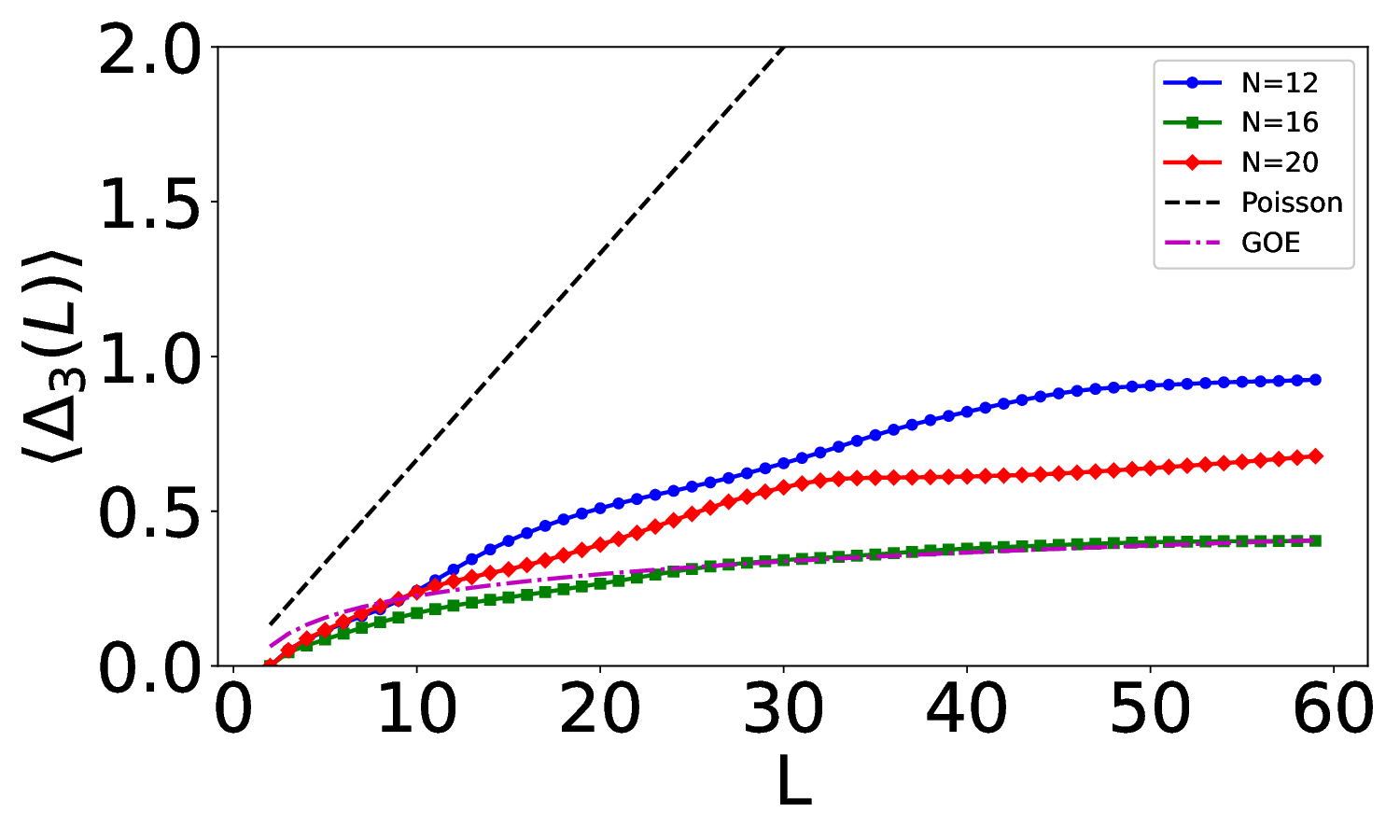}}
\subfigure[$L_{z}=N$, $g_{2}=0.3669$]{%
    \label{fig:rotvariancemoderate}
    \includegraphics[width=0.48\linewidth]{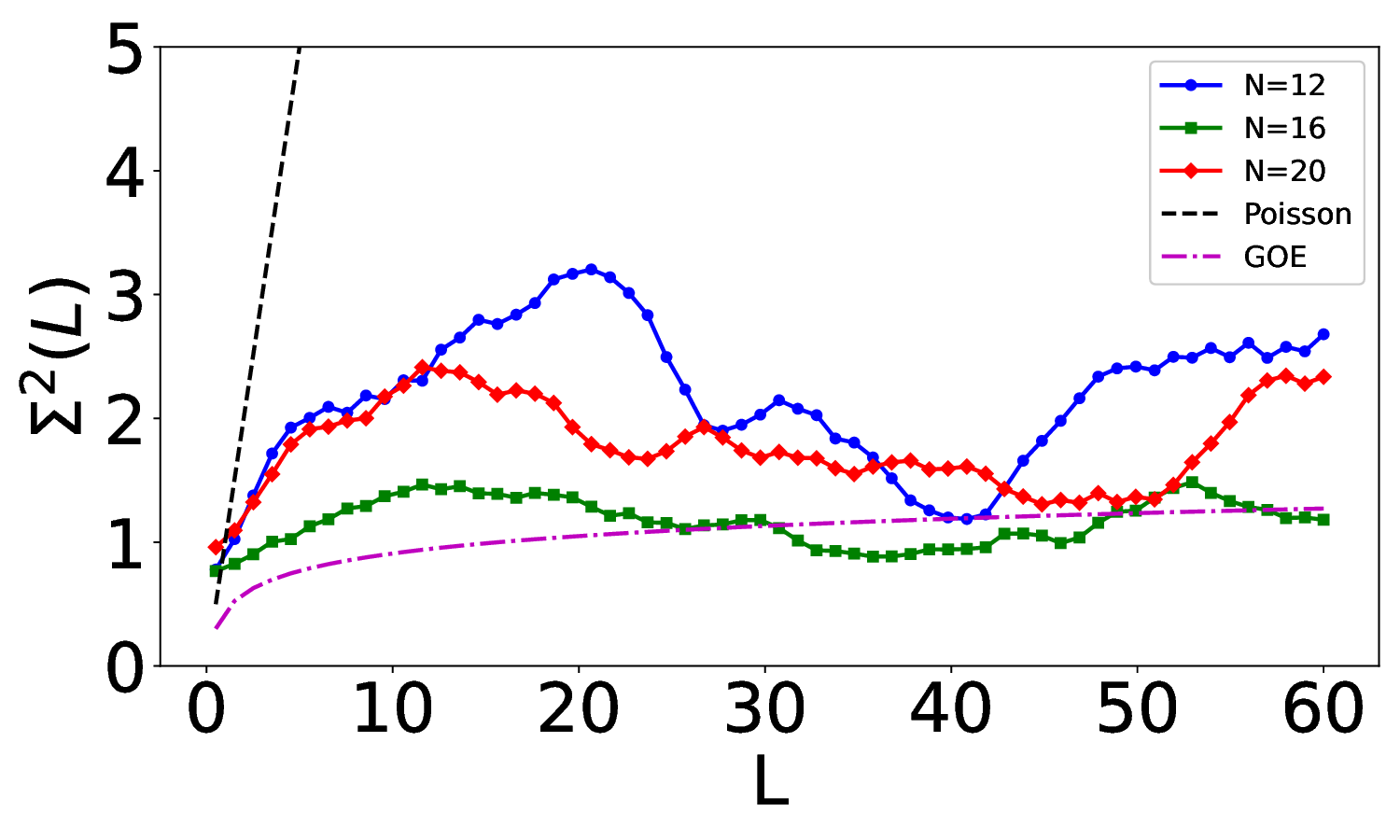}}

\subfigure[$L_{z}=N$, $g_{2}=3.669$]{%
    \label{fig:rotdeltastrong}
    \includegraphics[width=0.48\linewidth]{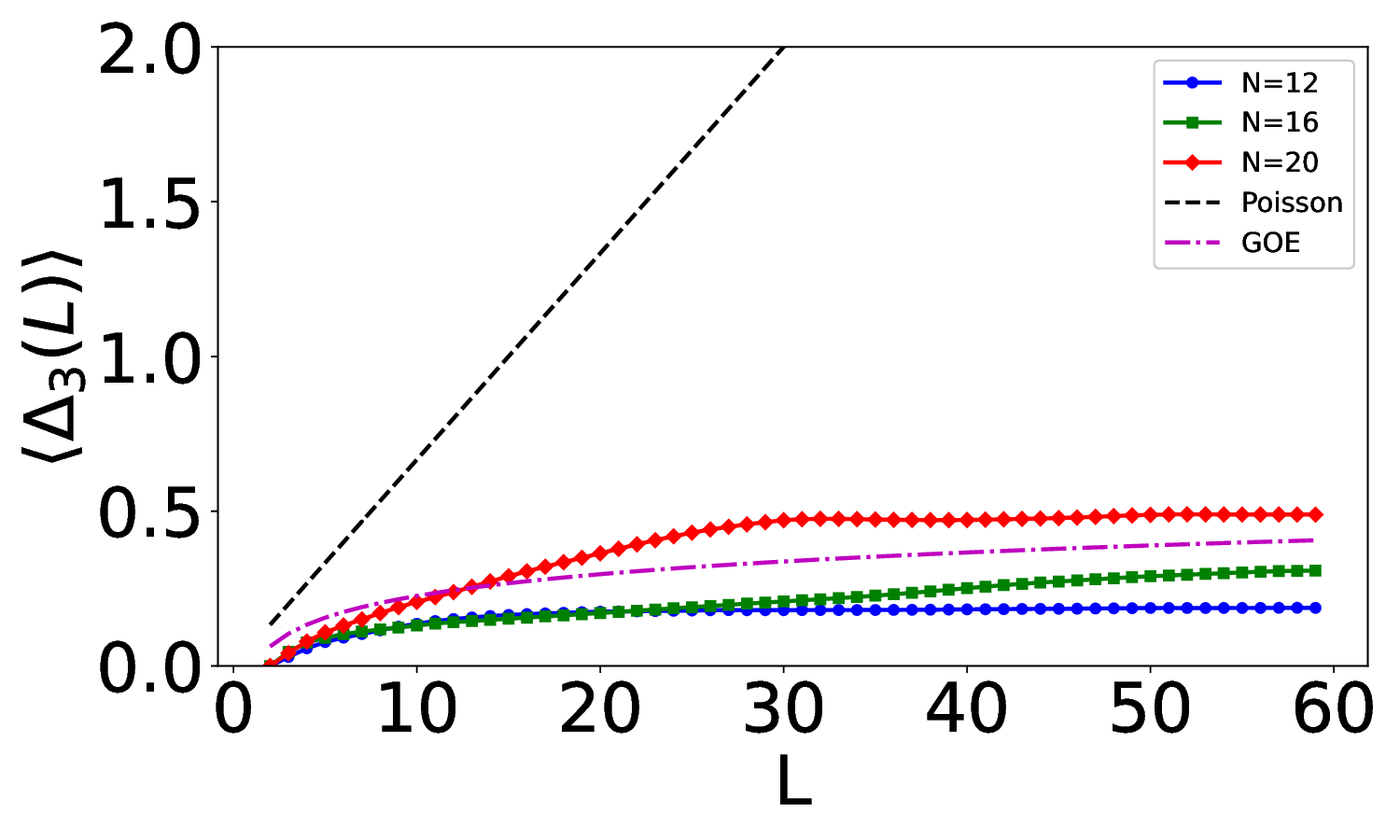}}
\subfigure[$L_{z}=N$, $g_{2}=3.669$]{%
    \label{fig:rotvariancestrong}
    \includegraphics[width=0.48\linewidth]{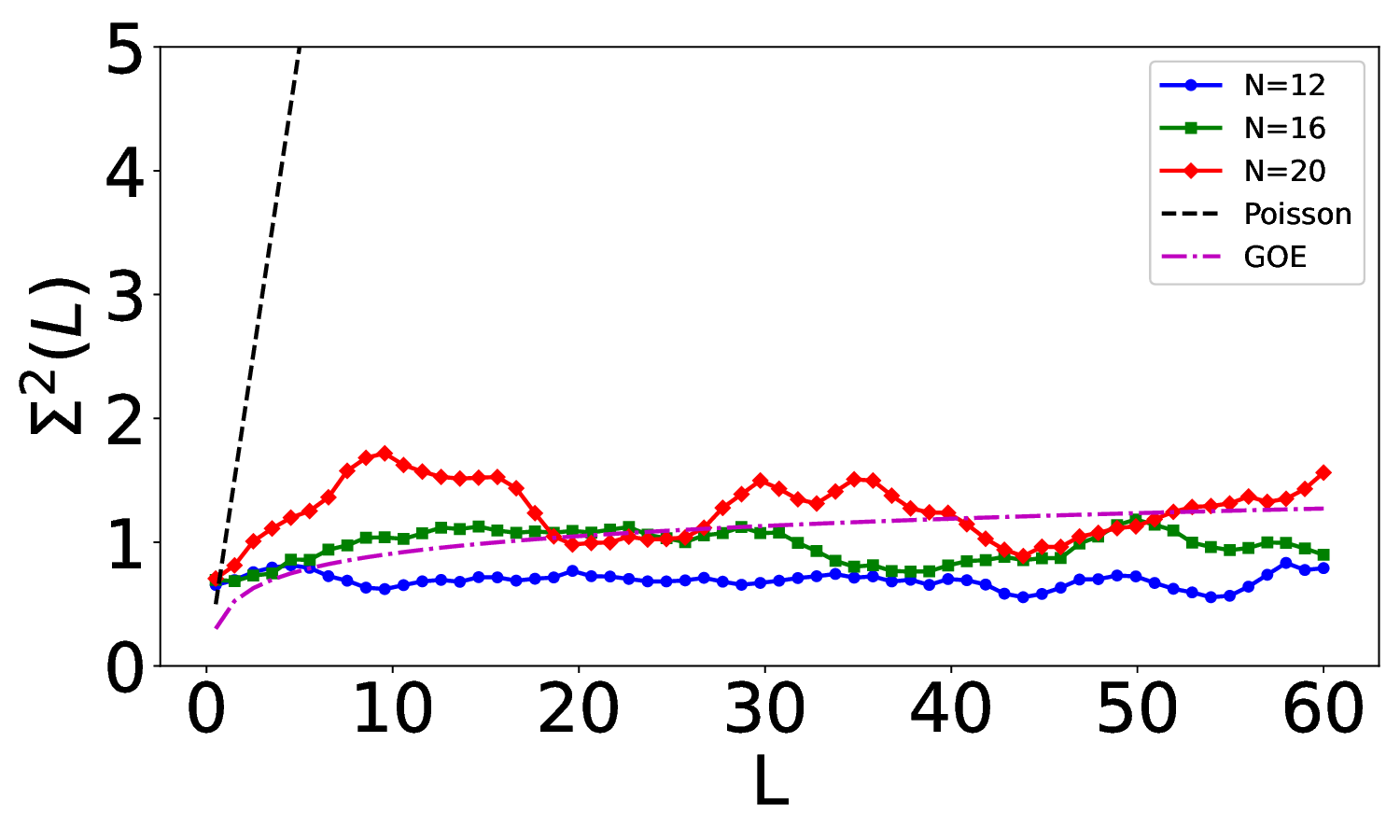}}

\vspace{-10pt}
\caption{(Color online) Spectral average $\langle \Delta_3(L) \rangle$ vs. $L$ and number variance $\Sigma^2(L)$ vs. $L$ for moderate interaction with $g_{2}=0.3669$ (upper panel) and strong interaction with $g_{2}=3.669$ (lower panel) regimes in the single-vortex state $L_z = N$ for $N = 12$, $16$, and $20$. Blue circles, green squares, and red diamonds show numerical results. Black dashed and magenta dash-dotted lines correspond to Poisson and GOE results, respectively.}
\label{fig:rotatinglongrange}
\end{figure*}

\begin{table*}[t]
\caption{The mean gap ratio with estimated error for the two subsets of energy levels—the lower 1–50 and the higher 51–100 levels—is calculated for different number of bosons in the rotating single-vortex case ($L_{z}=N$) in both moderate ($g_{2}=0.3669$) and strong ($g_{2}=3.669$) interaction regimes.}
\label{groupsgapratio2}
\centering
\begin{tabular}{|c|c|c|c|c|} 
\hline
 &Number of bosons, $N$ & Number of levels & \multicolumn{2}{c|}{Mean gap ratio $\langle r \rangle$} \\ 
\cline{4-5}
 & & & \textbf{$g_2 = 0.3669$} & \textbf{$g_2 = 3.669$} \\ 
\hline
 & \multirow{2}{*}{12} & 1--50 & 0.518 $\pm$ 0.041  & 0.569 $\pm$ 0.035 \\ 
\cline{3-5}
 &  & 51--100 & 0.478 $\pm$ 0.038 & 0.545 $\pm$ 0.038 \\ 
\hline
 & \multirow{2}{*}{16} & 1--50 & 0.529 $\pm$ 0.039 & 0.520 $\pm$ 0.037 \\ 
\cline{3-5}
 &  & 51--100 & 0.472 $\pm$ 0.040 & 0.486 $\pm$ 0.041 \\ 
\hline
 & \multirow{2}{*}{20} & 1--50 & 0.473 $\pm$ 0.035 & 0.564 $\pm$ 0.035 \\ 
\cline{3-5}
 &  & 51--100 & 0.517 $\pm$ 0.036 & 0.545 $\pm$ 0.030 \\ 
\hline
Poisson &  &  & \multicolumn{2}{c|}{0.386} \\ 
\hline
GOE     &  &  & \multicolumn{2}{c|}{0.530} \\ 
\hline
\end{tabular}
\end{table*}

\begin{figure*}[t]
    \centering  
         \subfigure[$N=12$, $L_{z}=2N$]{\label{fig:12brodyrotating}\includegraphics[width=0.32\linewidth]{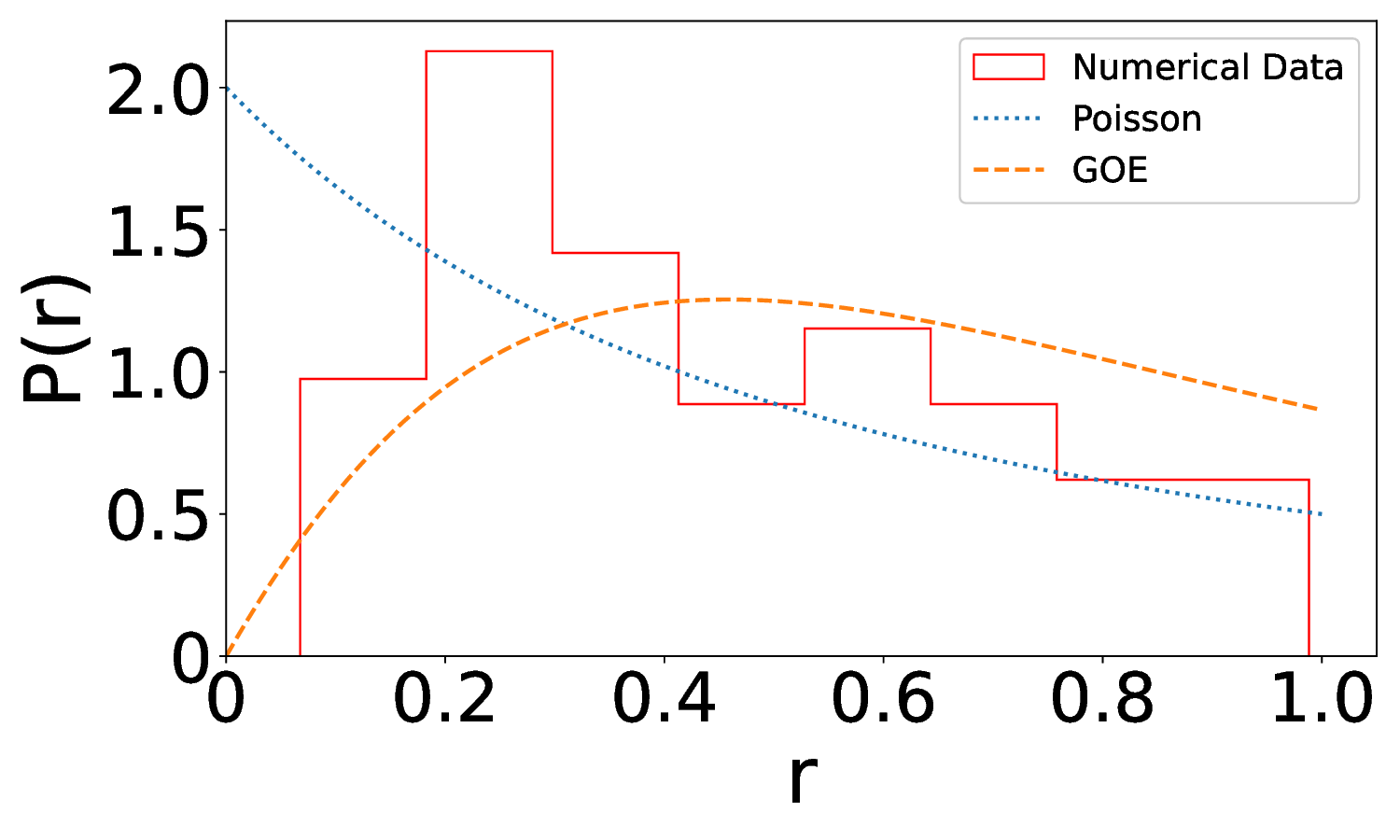}}
        \subfigure[$N=16$, $L_{z}=2N$]{\label{fig:24brodyrotating}\includegraphics[width=0.32\linewidth]{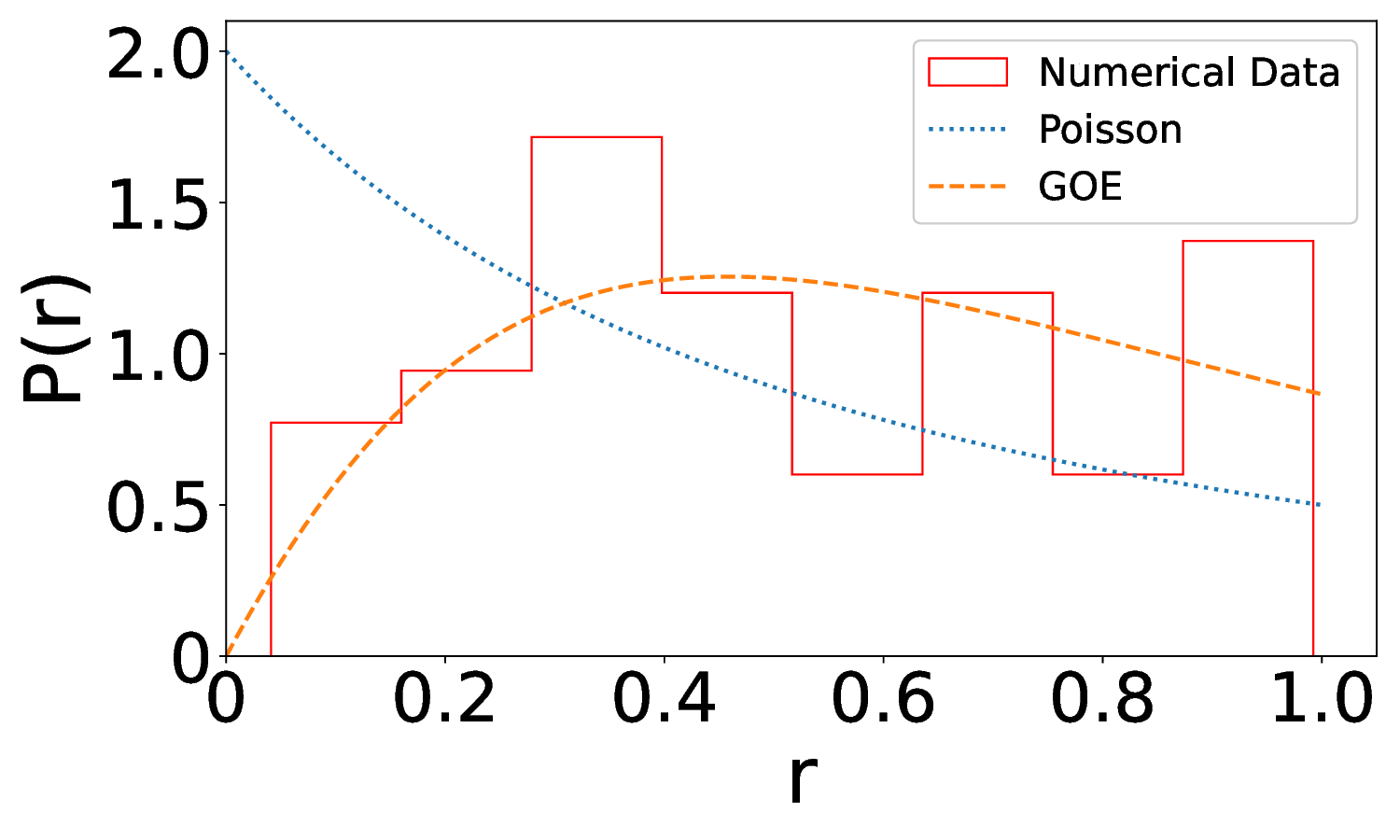}}
        \subfigure[$N=20$, $L_{z}=2N$]{\label{fig:36brodyrotating}\includegraphics[width=0.32\linewidth]{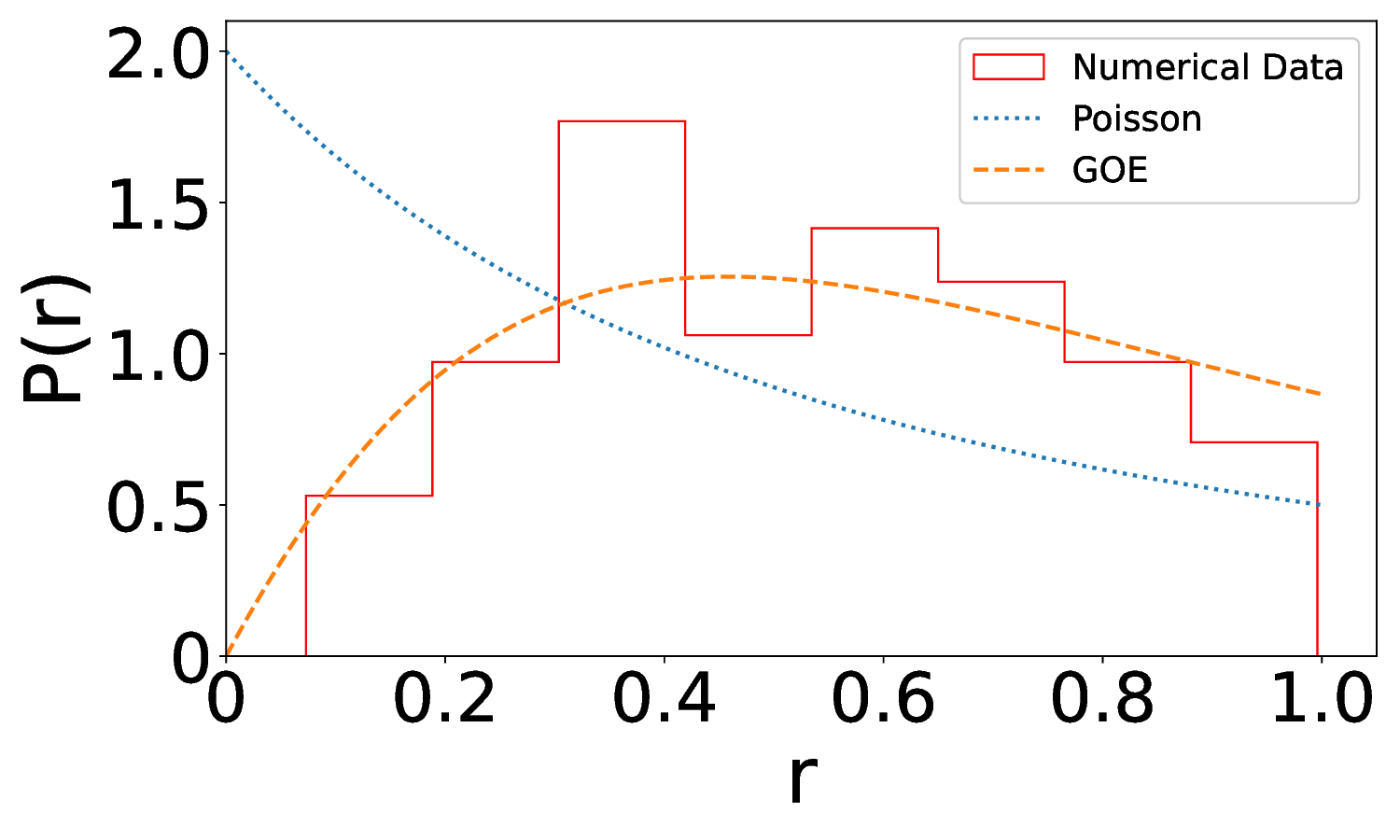}}
        \subfigure[$N=12$, $L_{z}=3N$]{\label{fig:12gaprotating}\includegraphics[width=0.32\linewidth]{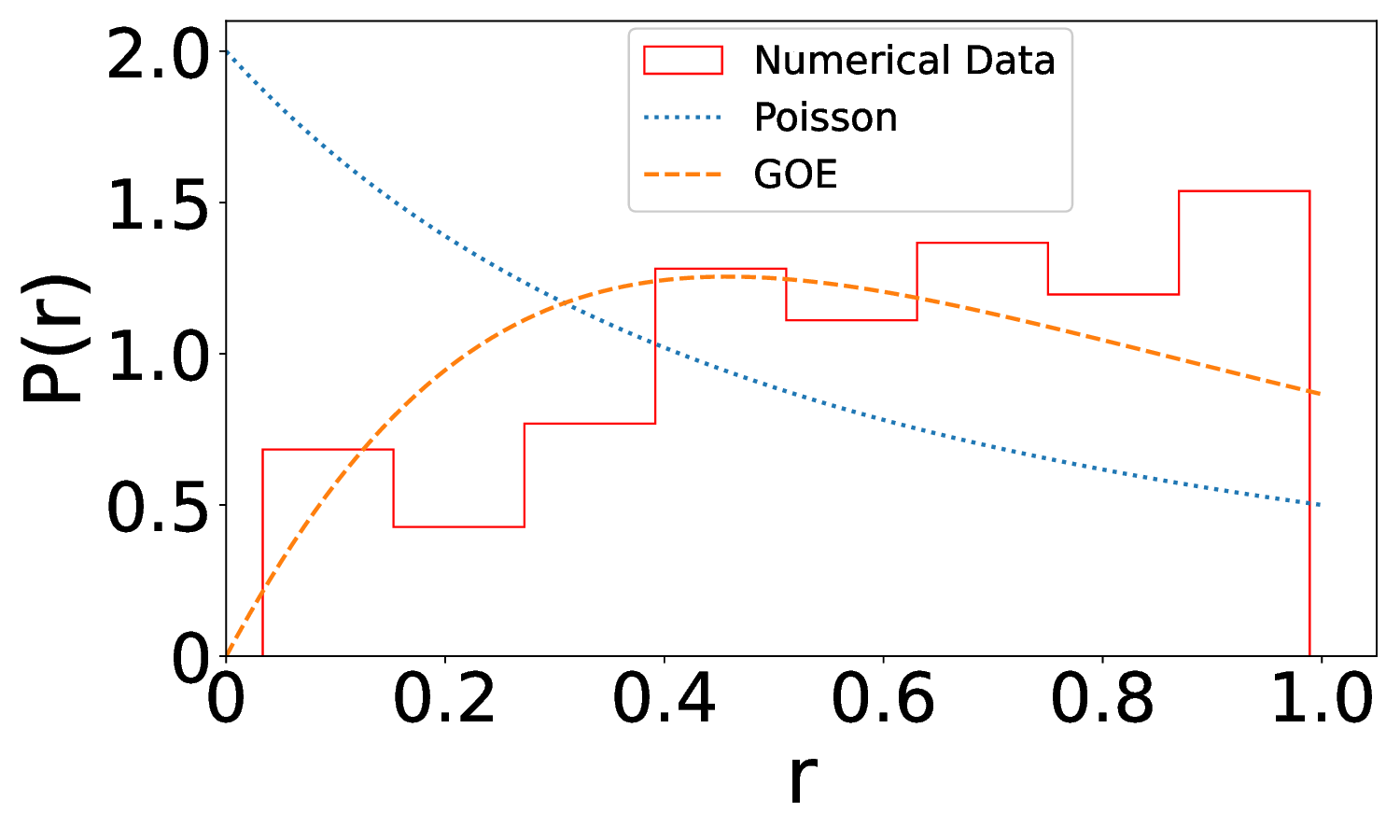}}
        \subfigure[$N=16$, $L_{z}=3N$]{\label{fig:24gaprotating}\includegraphics[width=0.32\linewidth]{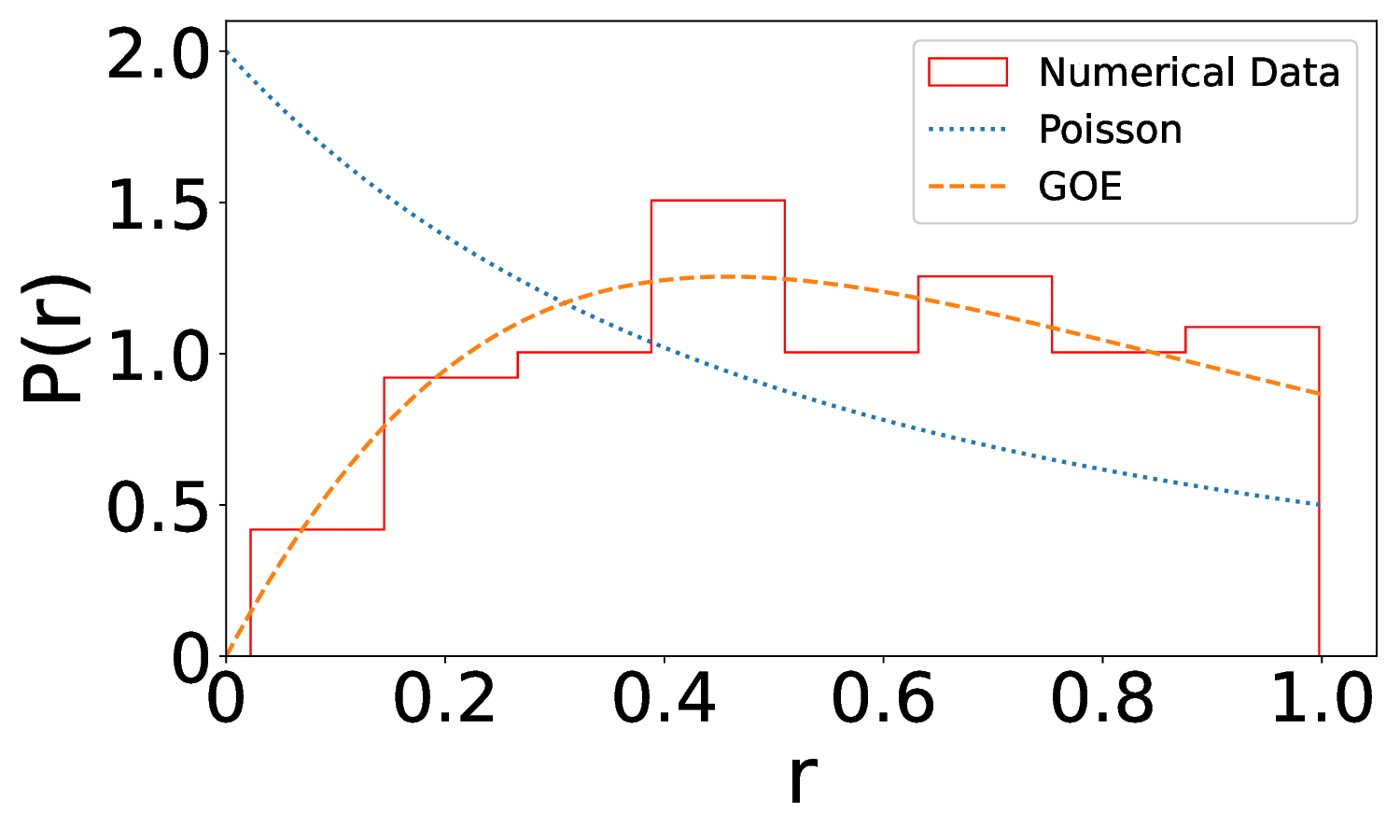}}
        \subfigure[$N=20$, $L_{z}=3N$]{\label{fig:36gaprotating}\includegraphics[width=0.32\linewidth]{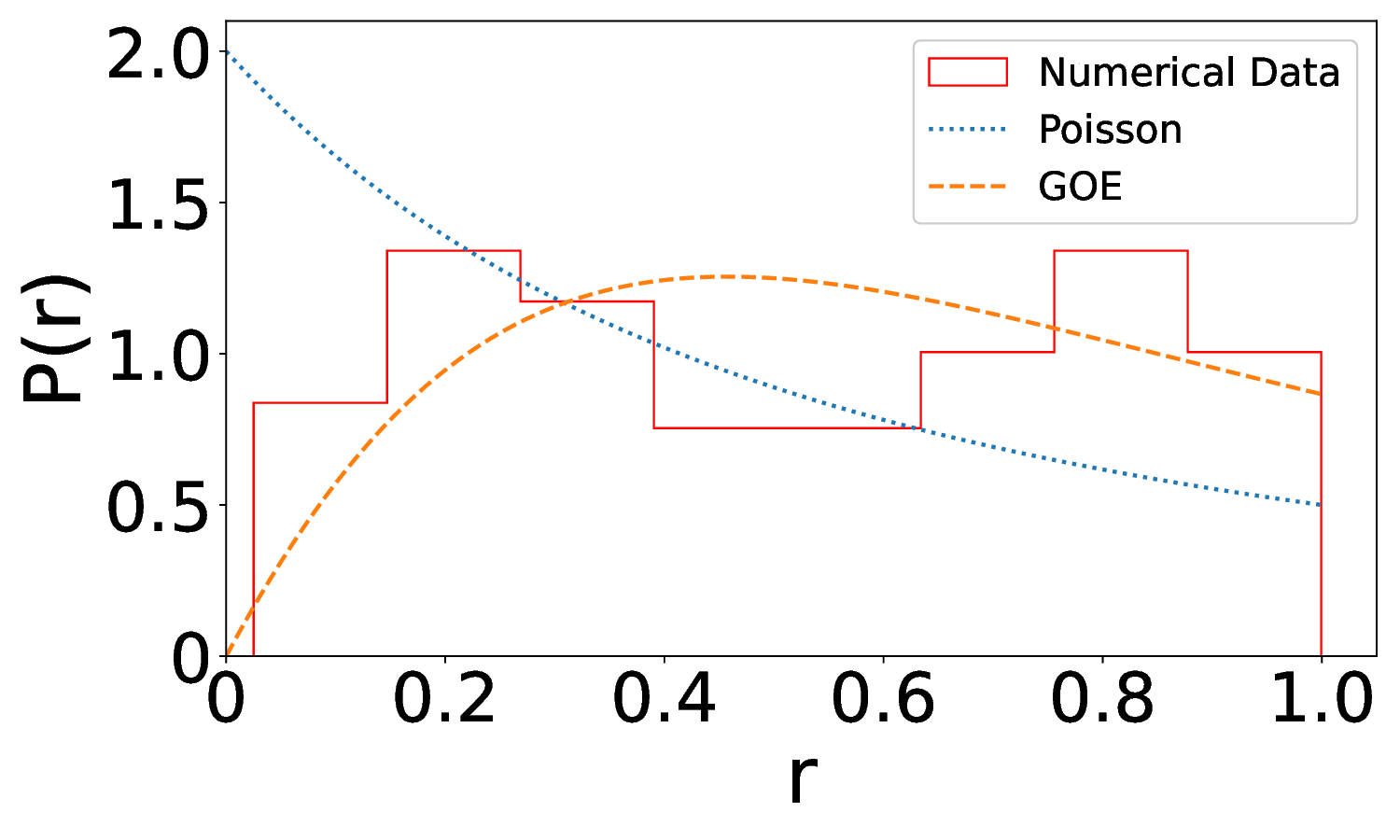}}
 \caption{(Color online) The distribution of the ratio of consecutive level spacings $P(r)$ for strong interaction regime with $g_{2}=3.669$ for $N=12, 16, 20$ in $L_{z}=2N$ (Upper Panel) and $L_{z}=3N$ (Lower Panel). The histogram in each graphs represents our numerical result for the lowest 100 energy levels. The blue line corresponds to the Poisson distribution, the orange dashed curve to the GOE distribution.}
    \label{fig:rotatingcaseLz=2N,3N}
\end{figure*}

\begin{figure*}[t]
    \centering 
      \subfigure[$L_{z}=2N$, $g_{2}=3.669$]{\label{fig:rotdeltastrongLz2N}\includegraphics[width=0.48\linewidth]{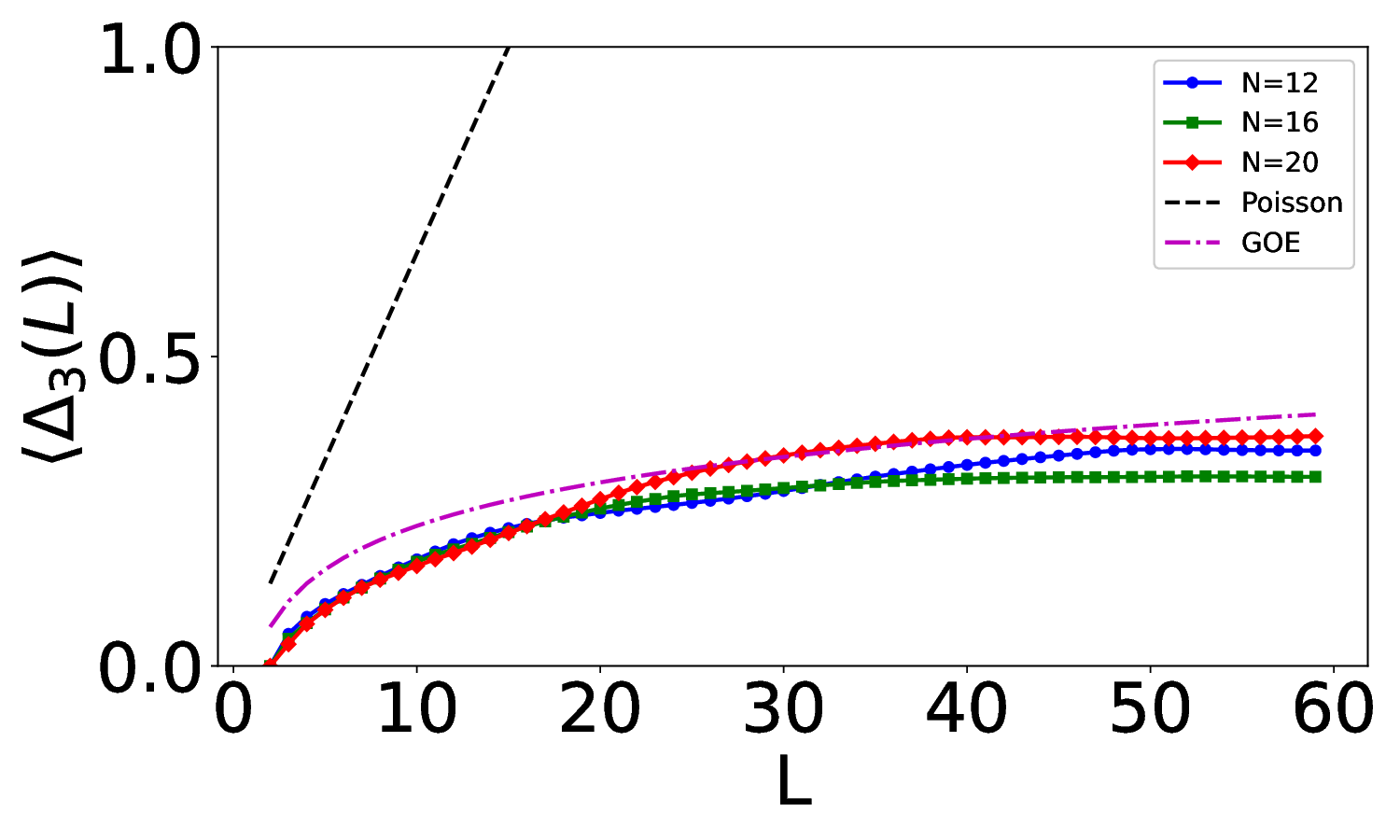}}
      \subfigure[$L_{z}=3N$, $g_{2}=3.669$]{\label{fig:rotdeltastrongLz3N}\includegraphics[width=0.48\linewidth]{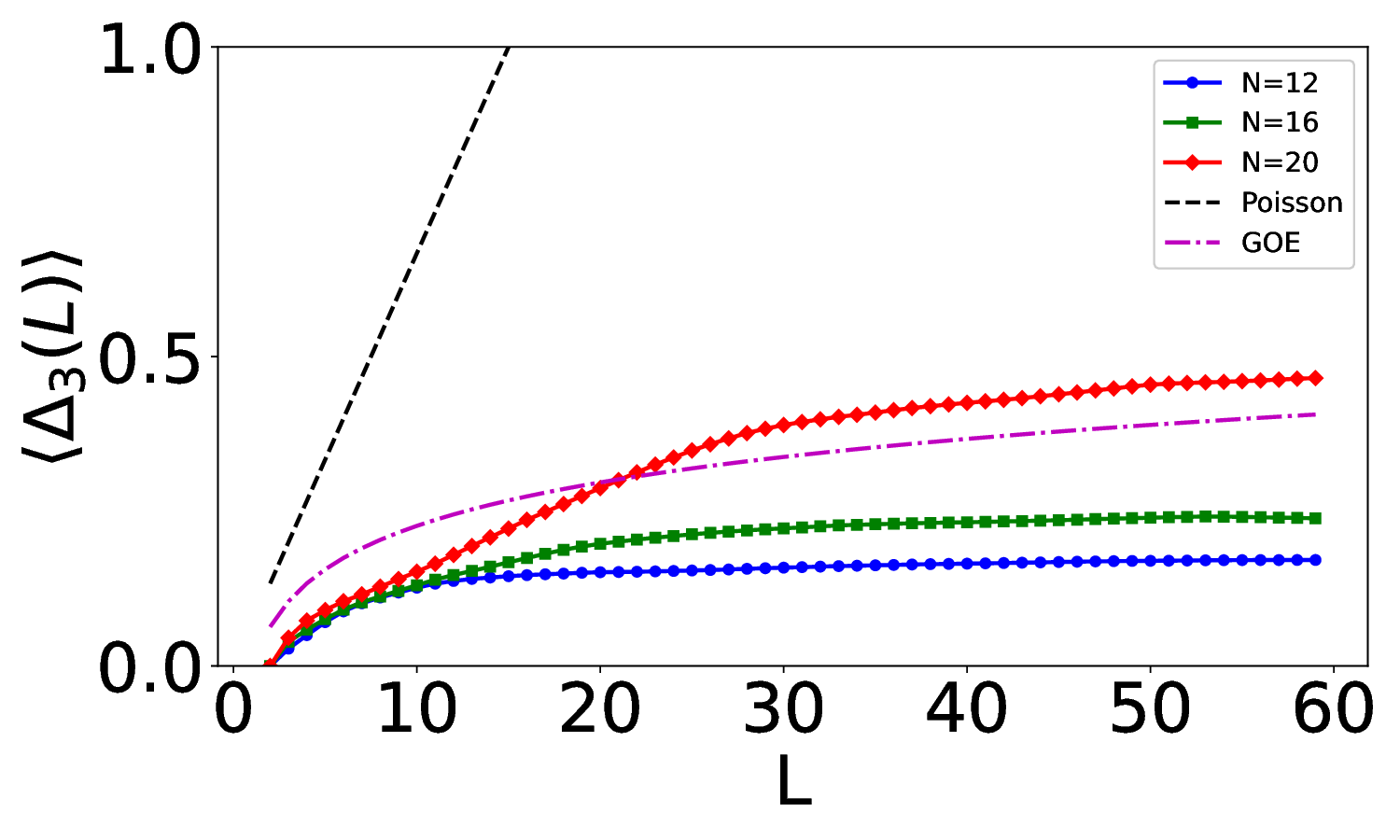}}

    \caption{(Color online) The spectral average $\langle \Delta_3(L) \rangle$ {\it vs} $L$ is presented for strong interaction regime with $g_{2}=3.669$ for different number of bosons $N=12, 16$ and $20$ in $L_{z}=2N$ (Left) and $L_{z}=3N$ (Right). The blue circle, green square and red diamond lines are our numerical results of $\langle \Delta_3(L) \rangle$ for $N=12,16$ and $20$, respectively, for the lowest 100 energy levels . For reference, we have also drawn $\langle \Delta_3(L) \rangle$ as a function of $L$, corresponding to the Poisson distribution (black dashed line) and the GOE distribution (magenta dash-dot line).}
    \label{fig:rotdeltastrongLz2N3N}
\end{figure*}

\subsubsection{\bf Short-range correlations}

The distributions $P(s)$ and $P(r)$ for various values of $L_z=N$ corresponding to number of bosons $N=12,16$ and $20$, are presented in Fig. \ref{fig:rotatingcasemoderate} for the moderate interaction regime and in Fig. \ref{fig:gaprot7.301} for the strong interaction regime.\\

\subsubsubsection{Moderate interaction regime}

For number of bosons $N=12, 16$, $20$ and for the single vortex state $L_{z}=N$, the NNSD $P(s)$ distribution exhibits GOE behavior with Brody parameter $b=0.52, 0.64$ and $0.64$, respectively, as shown in Figs. \ref{fig:12brodyrotatingcase}-\ref{fig:36brodyrotatingcase}. This indicates that these systems exist in a weakly chaotic regime, retaining substantial degree of regularity. 

The gap ratio distribution $P(r)$ for $N=12, 16$ and $20$ in the single vortex state $L_{z}=N$ are presented in Figs. \ref{fig:12gaprotatingcase}-\ref{fig:36gaprotatingcase} with mean gap ratio $\langle r \rangle = 0.498, 0.494$ and $0.496$, respectively, indicating that these systems remain in a weakly chaotic regime. The mean gap ratio corresponding to the lower ($1$–$50$) and the  higher ($51$–$100$) energy levels of the spectrum is listed in the fourth column of Table \ref{groupsgapratio2}. The lower part of the spectrum exhibits chaotic behavior consistent with GOE distribution ($\langle r \rangle \approx 0.530$), while the higher energy levels display weaker chaoticity, for $N=12$ and $16$. However, for $N=20$, the upper part (51-100) of the spectrum exhibits chaotic behavior consistent with GOE, while the lower part (1-50) is two standard deviations below GOE.\\

\subsubsubsection{Strong interaction regime}

The NNSD $P(s)$ for $N=12, 16, 20$ in the single vortex state $L_{z}=N$ is observed to follow the GOE distribution with Brody parameter $b=1.18, 0.98, 1.07$, respectively, as shown in Figs. \ref{fig:12nnsdrot7.301}-\ref{fig:36nnsdrot7.301}. In these figures, the peak of the histogram occurs near energy level spacing $s\approx 1$, indicating a significant accumulation of energy levels around $s\approx 1$. 

The gap ratio distribution $P(r)$ for $N=12, 16, 20$ in the single vortex state $L_{z}=N$ follows the GOE distribution with mean gap ratio $\langle r \rangle = 0.555, 0.505$ and $0.549$, respectively, as shown in Figs. \ref{fig:12gaprot7.301}-\ref{fig:36gaprot7.301}. This indicate that these systems show strong signature of quantum chaos. 

The Brody parameter $b$ and the mean gap ratio $\langle r \rangle$  values for each of the number of bosons $N$ in the single vortex state $L_{z}=N$ with moderate and strong interaction regimes are listed in Table \ref{tablebrodyrot2} and \ref{tablegapratiorot2}. The mean gap ratio for the lower ($1$–$50$) and higher ($51$–$100$) subsets of energy levels is presented in the fifth column of Table \ref{groupsgapratio2}. Both the lower and the higher parts of the spectrum exhibit chaotic behavior consistent with the GOE distribution ($\langle r \rangle \approx 0.530$).

\subsubsection{\bf Long-range correlations}

We now present our results on the Dyson-Mehta $\Delta_3(L)$ statistic and the level number variance $\Sigma^2(L)$ in both the moderate and the strong interaction regimes for the rotating single-vortex state $L_{z}=N$.\\

\subsubsubsection{Moderate interaction regime}

The spectral average $\Delta_3(L)$ for $N=12, 16, 20$ in the single vortex state $L_{z}=N$ is shown in Fig. \ref{fig:rotdeltamoderate}. It is observed that for small energy intervals $L$, the $\Delta_3(L)$ statistic exhibits GOE like behavior, while for large $L$, it approaches a saturation value. The saturated values for $N=12, 16$ and $20$ are found to be $\Delta_{\infty}$= $0.916$, $0.403$ and $0.657$, respectively. The value of $\Delta_{\infty}$ is modulated by the number of bosons $N$ in the moderate interaction regime which may be attributed to the interplay between the number of bosons and the interaction strength. 

The level number variance $\Sigma^2(L)$ for $N=12, 16, 20$ for the single vortex state $L_{z}=N$ is shown in Fig. \ref{fig:rotvariancemoderate}. We observed that for small values of $L$, $\Sigma^2(L)$ follows GOE distribution and deviates at large values of $L$. For $N=16$, the numerical data for $\Sigma^2(L)$ stays below the GOE distribution at small values of $L$, indicating further that the $N=16$ system exhibits the highest degree of chaos among the three systems studied.\\ 

\subsubsubsection{Strong interaction regime}

For small values of energy interval $L$, the numerical data of $\Delta_3(L)$ statistic for $N=12,16,20$ in the single vortex state $L_z = N$ follows the GOE distribution, as shown in Fig. \ref{fig:rotdeltastrong}. At large value of $L$, it saturates to a value $\Delta_{\infty}=0.187, 0.300$ and $0.489$ for $N=12, 16, 20$, respectively. These values are smaller than in the case of moderate interaction case $\Delta_{\infty}=0.916, 0.403, 0.657$ for the same number of bosons. This shows that the rotating system with strong interaction is strongly chaotic.

The level number variance $\Sigma^2(L)$ for $N=12,16,20$ in the single vortex state $L_{z}=N$ is shown in Fig. \ref{fig:rotvariancestrong}. For small values of $L$, the numerical data of $\Sigma^2(L)$ aligns with the GOE distribution, as shown in Fig. \ref{fig:rotvariancestrong}. At large values of $L$, the value of $\Sigma^2(L)$ deviates from the GOE distribution. However, the deviation from GOE behavior at large value of $L$ is smaller compared to the moderate interacting case, signifying that the strongly interacting rotating system exhibits strong signatures of chaos.

We also present the gap ratio distribution $P(r)$ and the spectral average $\Delta_3(L)$ for strong interaction regime ($g_{2}=3.669$) corresponding to angular momentum $L_{z}=2N$ and $L_{z}=3N$, as shown in Figs. \ref{fig:rotatingcaseLz=2N,3N} and \ref{fig:rotdeltastrongLz2N3N}, respectively. In Fig. \ref{fig:rotatingcaseLz=2N,3N}, we observe that the short-range correlation as measured by $P(r)$ follows the GOE distribution for both $L_{z}=2N$ (upper panel) and $L_{z}=3N$ (lower panel). Further, it is observed from Fig. \ref{fig:rotdeltastrongLz2N3N} that $\Delta_3(L)$ as a measure of the long-range correlation exhibits good agreement with the GOE distribution both for $L_{z}=2N$ (left) and $L_{z}=3N$ (right). Thus, the analysis of both short-range and long-range correlations exhibits signatures of quantum chaos for $L_{z}=2N, 3N$. This may be attributed to the interplay between strong two-body interaction, rotation and the number of bosons.

\section{\label{5} Conclusion}
In summary, we employed energy-level statistics to investigate the spectral properties of bosons harmonically trapped in a quasi-2D plane and interacting via repulsive Gaussian potential. We used the nearest-neighbor spacing distribution $P(s)$ and the distribution $P(r)$ of the ratio of consecutive level spacings for the short-range correlations while the Dyson-Mehta $\Delta_3$ statistic and the level number variance $\Sigma^2(L)$ were used for the long-range correlations. We considered both the situations i.e. when the interaction energy is small compared to the trap energy (moderate interaction regime) and when the interaction energy is comparable to the trap energy (strong interaction regime) for non-rotating as well as rotating cases. 
 
We observed that the non-rotating ($L_{z}=0$) system in the moderate interaction regime shows Poisson distribution, implying a regular behavior of the energy-level spectra. This is indicated by the NNSD and the gap-ratio distribution for the short-range correlations. The Dyson-Mehta $\Delta_3(L)$ statistic and the level number variance $\Sigma^2(L)$ follows Poisson distribution for small values of energy-interval $L$ for the long-range correlations. For larger values of $L$, the  $\Delta_3(L)$ statistic deviates from Poisson distribution  and saturates to a constant value. As the interaction increases to strong regime, the NNSD, the gap-ratio distribution, the $\Delta_3(L)$ statistic and the $\Sigma^2(L)$ align with the GOE distribution, implying a chaotic behavior with the degree of chaos modulated by the number of bosons $N$. It is observed that the $\Delta_3(L)$ statistic saturates to a much lower value in the strong interaction regime compared to the moderate interaction regime.

In the rotating case, for the single-vortex state $L_{z}=N$ in the moderate interaction regime, the system exhibits signatures of weak chaos with some degree of regularity, as indicated by the NNSD and the gap-ratio distribution. For small values of energy-interval $L$, the Dyson-Mehta $\Delta_3(L)$ statistic and the level number variance $\Sigma^2(L)$ follows the GOE distribution and for large values of $L$, the $\Delta_3(L)$ statistic saturates to a constant value, smaller than the non-rotating case.  As the interaction increases to strong regime, the NNSD, the gap-ratio distribution, the $\Delta_3(L)$ statistic and the $\Sigma^2(L)$ follows the GOE distribution, indicative of chaotic behavior. It is further observed that for $L_{z}=2N$ and $L_{z}=3N$, the system displays strong signatures of quantum chaos. 

Thus, the interplay between the interaction energy and the trap energy governs the transition of the system from regular to chaotic behavior, while rotation further amplifies the degree of chaos. Our study on trapped interacting bosons signifies the universal applicability of RMT in describing the spectral correlations in quantum many-body systems. We believe that these findings will contribute to advance the understanding of quantum chaos in ultracold Bose systems.

The present work may be readily extended to explore the spectral form factor to investigate the late time behavior of the trapped interacting Bose system. It would also be worthwhile to investigate how the energy-level statistics of an extended Bose-Hubbard model \cite{Benjamin2018} evolve as the strength of the cavity-induced infinite-range interactions is increased.

\begin{acknowledgments}
Mohd Talib thanks the Ministry of Social Justice and Empowerment, Government of India, for providing the Junior Research Fellowship (JRF) (\it NBCFDC Ref. No.: 221610104398).
\end{acknowledgments}

%\nocite{*}

%\bibliography{spectralstats}

\end{document}